\documentclass[%
reprint,
superscriptaddress,
preprintnumbers,
onecolumn,
 amsmath,amssymb,
 aps,showpacs,showkeys
]{revtex4-1}
\usepackage{graphicx}
\usepackage{xcolor}             
\usepackage{lineno}             
\usepackage{hyperref}           
\usepackage{setspace}           
\hypersetup{
    colorlinks=true,
    linkcolor=blue,
    filecolor=magenta,      
    urlcolor=cyan,
}
\begin{document}
\title{Critical parameters of the synchronisation's stability for coupled maps in regular graphs}

\author{Juan Gancio and Nicol{\'a}s Rubido}
\address{Universidad de la Rep{\'u}blica, Instituto de F{\'i}sica de Facultad de Ciencias, Igu{\'a} 4225, Montevideo 11400, Uruguay.}
\address{University of Aberdeen, King's College, Institute for Complex Systems and Mathematical Biology, AB24 3UE Aberdeen, United Kingdom.}


\begin{abstract}
Coupled Map Lattice (CML) models are particularly suitable to study spatially extended behaviours, such as wave-like patterns, spatio-temporal chaos, and synchronisation. Complete synchronisation in CMLs emerges when all maps have their state variables with equal magnitude, forming a spatially-uniform pattern that evolves in time. Here, we derive critical values for the parameters -- coupling strength, maximum Lyapunov exponent, and link density -- that control the synchronisation-manifold's linear stability of diffusively-coupled, identical, chaotic maps in generic regular graphs (i.e., graphs with uniform node degrees) and class-specific cyclic graphs (i.e., periodic lattices with cyclical node permutation symmetries). Our derivations are based on the Laplacian matrix eigenvalues, where we give closed-form expressions for the smallest non-zero eigenvalue and largest eigenvalue of regular graphs and show that these graphs can be classified into two sets according to a topological condition (derived from the stability analysis). We also make derivations for two classes of cyclic graph: $k$-cycles (i.e., regular lattices of even degree $k$, which can be embedded in $T^k$ tori) and $k$-M{\"o}bius ladders, which we introduce here to generalise the M{\"o}bius ladder of degree $k = 3$. Our results highlight differences in the synchronisation manifold's stability of these graphs -- even for identical node degrees -- in the finite size and infinite size limit.
\end{abstract}
\date{\today}
\maketitle
\section*{Introduction}\label{sec_Intro}
Coupled Map Lattices (CMLs) were introduced as suitable models to study the behaviour of spatially extended dynamical systems \cite{Kaneko_1984}. These systems are defined on a discrete space-time, but with state variables that can take continuous values. Their behaviours range from ordered wave-like patterns to spatio-temporal chaos (i.e., turbulence) \cite{Kaneko_1989,Amritkar_1993,Gallas_2004,Lakshmanan_2005}. CMLs have also been generalised to include non-local interactions \cite{Kaneko_1990}, either by using distance-dependent functions \cite{Viana_2003,Batista_2005,Rubido_2011} or by replacing the lattice regularity with complex graphs \cite{Strogatz_1998,Sync_SW_barahona_pecora,Strogatz_2006,Boccaletti_2005}, and delayed interactions \cite{Masoller_2005,Masoller_2006,Masoller_2009}. Overall, CMLs (and their generalisations) have allowed to deepen our understanding of complex behaviours, such as intermittence \cite{Kaneko_1985,Hilda_1996}, chimera states \cite{Omelchenko_2011,Scholl_2012,Baptista_2020}, and synchronisation \cite{Boccaletti_1998,Gade_2000,Jost_2001,Boccaletti_2002}.

Complete synchronisation (CS) is one of these collective behaviours emerging in many natural systems and with broad real-world applications, such as the design of stable power-grids \cite{Dorfler_2013,Nardelli_2014,Motter_2015}. For CMLs, CS implies having all maps evolving such that their state variables have identical values at any time; that is, a spatially-uniform pattern. The evolution and stability of this pattern can be analysed, for example, by means of Lyapunov exponents \cite{Kaneko_1986,Viana_2007}, which are related to the Kaplan-Yorke dimension \cite{Yorke_1979,Yorke_1983} and Kolmogorov-Sinai Entropy \cite{Kolmogorov_1958,Sinai_1959,Viana_2002} of the system. Research on synchronisation generally focuses on understanding which dynamical properties and topological characteristics favour -- or hinder -- the emergence of CS.

A major breakthrough in synchronisation research was achieved by Pecora and Carroll \cite{MSF_pecora}, whose seminal work defined the Master Stability Function (MSF): a functional analysis of the synchronisation manifold's stability for generic graphs of diffusively-coupled, identical, (time-continuous or discrete) dynamical systems. The MSF allows to decouple the dynamical properties of the dynamical units composing the coupled system with its topological properties (similar to the work by Fujisaka and Yamada \cite{Fujisaka_MSF}). In spite of the MSF breakthrough, and because of the broad range of dynamics and graphs that can be analysed, there are still plenty of open-questions that can aid in the design of stable synchronous systems and continue increasing our understanding of this fascinating collective phenomenon.

Here, we derive closed-form expressions for the minimum coupling strength and link density necessary to have a stable synchronisation -- as well as an upper limit to the chaoticity that can be synchronisable -- of diffusively-coupled, identical maps, in generic regular graphs and class-specific cyclic graphs. Cyclic graphs are lattices having cyclical node-permutation symmetry (implying periodic boundary conditions and identical node neighbourhoods). In particular, we make derivations for $k$-cycles (also known as Wiley-Strogatz-Girvan networks \cite{Strogatz_2006}) and $k$-M{\"o}bius ladders (non-planar graphs), which we introduce in this work to extend the classic M{\"o}bius ladder with degree $3$ \cite{mob_lad_org,mob_lad} to higher degrees. Our finite-size results show striking differences between these $2$ cyclic graphs, only becoming similar when converging to the complete (all-to-all) graph. Moreover, we show that our expressions can change for different degrees and in the thermodynamic limit (i.e., infinite system size). Our derivations are based on the MSF \cite{MSF_pecora} and the graph's Laplacian eigenvalues (focusing on the smallest non-zero and largest eigenvalue), making our approach general. Overall, our work complements the general understanding of synchronisation phenomena in CMLs and provides detailed mathematical derivations leading to exact analytical results.
%
\section*{Methods and Model}\label{sec_Methods}

    \subsection*{Coupled Map Lattices and the Master Stability Function}
Let $N$ one-dimensional maps, $f_i:D\subset\mathbb{R} \to D$, where $i = 1,\ldots,N$ (corresponding to possibly different parameters), be diffusively coupled in a symmetric graph \cite{Kaneko_1984},
\begin{equation} \label{kaneko}
    x_{t+1}^{(i)} =  f_i\!\left(x_{t}^{(i)}\right) - \frac{\epsilon}{k_i} \sum_{j=1}^N L_{ij}\, f_j\!\left(x_{t}^{(j)}\right),
\end{equation}
where $0\leq\epsilon\leq1$ is the coupling strength and $L_{ij}$ is $ij$-th element of the graph's Laplacian matrix. $\mathbf{L} = \mathbf{K} - \mathbf{A}$, where $\mathbf{A}$ is the graph's adjacency matrix ($A_{ij} = 1 = A_{ji}$ if there is a link between nodes $i$ and $j$, and $A_{ij} = 0$ otherwise) and $k_i = \sum_{j=1}^N A_{ij}$ is the $i$-th degree (number of neighbours). Equation~(\ref{kaneko}) describes an $N$-dimensional mapping, that transforms the state of the $N$ maps at instant $t$, $\vec{x}_t = \{x_t^{(1)},\dots,x_t^{(N)}\}$, to the state, $\vec{x}_{t+1} = \{x_{t+1}^{(1)},\dots,x_{t+1}^{(N)}\}$. This mapping can be written in matrix form as
\begin{equation}\label{matrix_form}
   \vec{x}_{t+1} =  \left[ \mathbf{I} - \epsilon\,\mathbf{K}^{-1}\,\mathbf{L}\right]\,\vec{f}\!\left(\vec{x}_t\right), 
\end{equation}
where $\vec{f}\!\left(\vec{x}_t\right) = \{f_1(x_t^{(1)}),\ldots,f_N(x_t^{(N)})\}$ represents the mapping of each of the $N$ maps at time $t$, $\mathbf{I}$ is the $N \times N$ identity matrix, and $\mathbf{K}^{-1} = \textnormal{diag}\{1/k_1,\ldots,1/k_N\}$.

When the coupled system is composed of identical mappings, $f_i = f\;\forall\,i$, $s_t = x_t^{(1)} = \dots = x_t^{(N)}$ is a solution of Eq.~(\ref{matrix_form}) because of the zero-row-sum property of $\mathbf{L}$ (i.e., $\sum_j L_{ij} = 0\;\forall\,i$). This solution defines the complete synchronisation (CS) manifold, whose linear stability is determined by the Master Stability Function (MSF) \cite{MSF_pecora}. Specifically, the stability is quantified by the Lyapunov exponents transverse to the synchronisation manifold, which are known as Conditional Lyapunov Exponents (CLE), $\chi$, because their validity is restricted to the diagonal of the $N$-dimensional state-space.

In terms of the MSF, the system is able to synchronise if the transverse CLEs are negative; meaning that perturbations to the manifold decay exponentially fast and the manifold is linearly stable. This situation is generally possible if $\frac{\alpha_2}{\alpha_1} = \beta > \frac{\lambda_{M}}{\lambda_{F}}$, where $\lambda_{F}$ is the Fiedler's eigenvalue of $\mathbf{L}$ (i.e., the first non-zero eigenvalue), $\lambda_{M}$ is its largest eigenvalue, and $\alpha_1$ and $\alpha_2$ are the limits defining the negative range of CLEs \cite{Sync_SW_barahona_pecora}, which depend on the system's dynamical characteristics and coupling strength.

In particular, the MSF is obtained by perturbing the synchronous state and analysing the perturbation's evolution up to the leading order. In Eq.~(\ref{matrix_form}), such perturbation, $x_t^{(i)}=s_t+\xi_t^{(i)}$, up to the first order in $\xi_t^{(i)}$, holds
$$ \vec{\xi}_{t+1} =\left[ \mathbf{I} - \epsilon\,\mathbf{K}^{-1}\,\mathbf{L}\right] \mathbf{J}_{\!\vec{f}}(s_t)\,\vec{\xi}_{t},$$
where $\mathbf{J}_{\!\vec{f}}(s_t)$ represent the Jacobian matrix of $\vec{f}$ evaluated in the synchronous state $s_t$. In our case, $\mathbf{J}$ is a diagonal matrix -- even for non-synchronous solutions. Specifically, $\mathbf{J}_{\vec{f}}(\vec{x}_t) = \textnormal{diag} \{\partial_1 f_1(x_t^{(1)}),\ldots,\partial_N f_N(x_t^{(N)})\}$, with $\partial_i f_i = d\,f_i/d\,x^{(i)}$ being the derivatives of the flow-vector components with respect to each independent variable. Thus, when $f_i = f\;\forall\,i$, the synchronisation manifold Jacobian matrix is given by $\mathbf{J}_{\vec{f}}(s_t)=f'(s_t)\,\mathbf{I}$, which lead to
\begin{equation}
    \vec{\xi}_{t+1} =f'(s_t)\left[ \mathbf{I} - \epsilon\,\mathbf{K}^{-1}\,\mathbf{L}\right]\,\vec{\xi}_{t}.
    \label{eq_MSF_synch}
\end{equation}
This is a linear mapping done by a constant matrix, $\mathbf{I} - \epsilon\,\mathbf{K}^{-1}\,\mathbf{L}$, to the perturbations at time $t$, $\vec{\xi}_t$, modulated by the map's derivative at the synchronisation manifold, $f'(s_t)$.

    \subsection*{Synchronisation Stability in Regular and Cyclic Graphs}
We restrict our analysis of Eq.~(\ref{eq_MSF_synch}) to coupled-maps in regular graphs, such that $\mathbf{K}^{-1}=\frac{1}{k} \mathbf{I}$, which commutes with any matrix. We note that for symmetric graphs, $\mathbf{L}$ is Hermitian, meaning that it can be diagonalised and that it holds real eigenvalues. Thus, we write $\mathbf{L} = \mathbf{P} \mathbf{\Lambda} \mathbf{P}^{-1}$, where $\mathbf{\Lambda} = \textnormal{diag}\{\lambda_0,\ldots,\lambda_{N-1}\}$ is the ordered eigenvalue spectra (with $\lambda_0 = 0 < \lambda_1 \leq \cdots \leq \lambda_{N-1}$) and $\mathbf{P} = \{\vec{\psi}_0,\ldots,\vec{\psi}_{N-1}\}$ holds their respective orthonormal (column) eigenvectors, such that $\mathbf{L}\,\vec{\psi}_n = \lambda_n\,\vec{\psi}_n\;\forall\,n$. Consequently, changing variables in Eq.~(\ref{eq_MSF_synch}) to $\vec{\zeta}_{t} = \mathbf{P}^{-1}\vec{\xi}_{t}$, the perturbations to the synchronisation state become decoupled in the eigenmodes ($n = 0,1,\ldots,N-1$) according to
\begin{equation}\label{diag_pert}
        \zeta^{(n)}_{t+1} = \left( 1 - \epsilon\,\frac{\lambda_n}{k} \right) f'(s_t)\,\zeta^{(n)}_{t}.
\end{equation}

Equation~(\ref{diag_pert}) gives the system's CLEs, $\{\chi_n\}_{n=0}^{N-1}$ when iterated; that is \cite{Hilda_1996,Viana_2003,Jost_2001}, $ \chi_n = \log\left|1-\epsilon\,\frac{\lambda_n}{k}\right| + \lim\limits_{T \to \infty} \sum_{t=1}^T  \frac{\log \left|f'(x_t)\right|}{T} = \chi_{top}( \epsilon \lambda_n/k ) + \chi_{dyn} $, where $\chi_0 = \chi_{dyn}$ (because $\lambda_0 = 0$ always) is the exponent parallel to the synchronisation manifold, i.e., the isolated map's (constant) Lyapunov exponent, and the remaining $N-1$ exponents determine the stability of the manifold (transversal directions), being stable if $\chi_n < 0$ $\forall\,n>0$. This means that a stable manifold necessary has transversal modes fulfilling
\begin{equation}
    \chi_{top}(\epsilon\lambda_n/k) = \log\left|1-\epsilon\,\frac{\lambda_n}{k}\right| < - \chi_{dyn},\;\forall\,n>0.
    \label{eq_StabilityCond}
\end{equation}

We note that when $\chi_{dyn}\leq0$, Eq.~(\ref{eq_StabilityCond}) is always satisfied, meaning that periodic dynamics have linearly-stable synchronisations. On the other hand, when the map is sufficiently chaotic, $\chi_{dyn} \gg 0$, the negative well of the MSF can be narrowed down to the point of disappearing. Hence, the system's ability to synchronise depends on the competition between the map's chaoticity and the network's topology, which we explore in detail in this work focusing on chaotic maps, i.e., $\chi_{dyn}>0$.

Cyclic graphs are a particular class of regular graphs: they preserve their topology when transformed by a group of symmetries which cyclically takes any one node and maps it into another. Namely, a cyclic permutation is such that $\pi[\{1,2,\ldots,N-1,N\}] = \{2,\ldots,N-1,N,1\}$ (hence, $\pi\circ\pi\circ\cdots\circ\pi = \pi^N = \mathbf{I}$), and cyclic graphs are graphs that preserve their local and global topological properties under groups of permutations, $\{\pi,\pi^2,\ldots,\pi^N\}$. This implies that cyclic graphs contain all their connectivity information in any given row of $\mathbf{L}$ (or $\mathbf{A}$) and have analytical expressions for their eigenvalues and eigenvectors based on a Fourier basis \cite{CGraphs,GraphSpectra,FanChung}. Hence, we will focus on the first row of $\mathbf{L}$, $\{L_{1,j}\}_{j=1}^{N} = \{k,\,- A_{1,2},\dots,-A_{1,N}\}$, and the eigenvalues can be expressed in terms of $\{L_{1,j}\}$ as
\begin{equation}
    \lambda_n = \sum_{j=1}^{N} L_{1,j} \cos\!\left[\frac{2 \pi n}{N}(j-1)\right] = k - \sum_{j=2}^{N} A_{1,j} \cos\!\left[\frac{2 \pi n}{N}(j-1)\right].
    \label{eq:eigenvals}
\end{equation}

We note that from Eq.~(\ref{eq:eigenvals}) the eigenvalue-magnitudes are symmetric due to the cosine function, implying that $\lambda_n = \lambda_{N-n+1}\;\forall\,n>0$ and $\lambda_0 = 0$ for any cyclic graph. This implies that almost every eigenvalue is (at least) doubly degenerate, except for $\lambda_0 = 0$. Also, we note that the smallest non-zero eigenvalue, $\lambda_F$ (known as Fiedler eigenvalue \cite{Fiedler_1973} or algebraic connectivity), or the maximum eigenvalue, $\lambda_M$, of a given cyclic graph, can be different than $\lambda_1$ or $\lambda_{N/2}$ from Eq.~(\ref{eq:eigenvals}), respectively.

\section*{Results}\label{sec_Results}
We analyse diffusively-coupled, identical, chaotic maps in generic -- and specific -- regular graphs to find the necessary conditions to have a linearly stable synchronisation manifold. Our main contributions are the derivation of critical parameters, including eigenvalue magnitudes, minimum coupling strengths, map's maximum Lyapunov exponent (i.e., maximum synchronisable chaoticity), and link-density. In particular, we derive closed-form expressions for these critical parameters in $2$ specific classes of cyclic graphs for the finite and infinite size limits: $k$-cycles -- ring-like graphs connecting an even number of $k$ neighbours -- and $k$-M{\"o}bius ladders -- which we introduce to generalise the M{\"o}bius ladder (of degree $k = 3$) to $3 \leq k \leq N-1$. 

    \subsection*{Synchronisation-Manifold's Stability for Generic Regular Graphs}
The stability condition set by Eq.~(\ref{eq_StabilityCond}) depends on the map's Lyapunov exponent, $\chi_{dyn}$, Laplacian matrix's eigenvalues, $\{\lambda_n\}_{n=1}^{N-1}$, graph's degree, $k$, and coupling strength, $\epsilon$. Laplacian eigenvalues are such that $\{\lambda_n\}_{n=0}^{N-1} = \{ \lambda_0 = 0 < \lambda_F \leq \cdots \leq \lambda_M \}$, where $\lambda_F$ is the Fiedler eigenvalue \cite{Fiedler_1973}, i.e., smallest non-zero eigenvalue (also known as algebraic connectivity), and $\lambda_M$ is the largest eigenvalue. These $2$ are the relevant eigenvalues to analyse the synchronisation-manifold's stability \cite{MSF_pecora}. Hence, we rewrite the condition set by Eq.~(\ref{eq_StabilityCond}) in terms of bounds to $\lambda_F$ and $\lambda_M$ \cite{Jost_2001} by
\begin{equation}\label{eq:stability_cond}
     \mathcal{S}_F(\chi_{dyn},\,\epsilon) < \frac{\lambda_F}{k} \leq \frac{\lambda_M}{k} < \mathcal{S}_M(\chi_{dyn},\,\epsilon),
\end{equation}
where $\mathcal{S}_F(\chi_{dyn},\,\epsilon) \equiv [1 - \exp(\chi_{dyn})]/\epsilon$ and $\mathcal{S}_M(\chi_{dyn},\,\epsilon) \equiv [1 + \exp(\chi_{dyn})]/\epsilon$ define $2$ non-intersecting surfaces, with $\mathcal{S}_M(\chi_{dyn},\,\epsilon) > 1,\;\forall\,\chi_{dyn} > 0$ and $\epsilon\in(0,\,1]$. The inequalities in Eq.~(\ref{eq:stability_cond}) determine a lower and an upper bound for $\lambda_F/k$ and $\lambda_M/k$ as a function of $\epsilon$ and $\chi_{dyn}$, such that when fulfilled, all transversal directions to the synchronisation manifold are attractive and the system has a linearly stable synchronisation.

We note that, from the Gershgorin's Circle theorem \cite{Gershgorin_1931}, all Laplacian eigenvalues are bounded to the interval $[0,\,2k_M]$, where $k_M = \max\{k_i\}_{i=1}^N$. For regular graphs, this implies that $\lambda_n/k \in (0,\,2]\;\forall\,n>0$, meaning that Eq.~(\ref{eq:stability_cond}) restricts the interval $[\lambda_F,\lambda_M]\subset(0,2]$ between the surfaces -- stability is lost whenever this eigenvalue interval intersects a surface. In what follows, we use Eq.~(\ref{eq:stability_cond}) to determine the critical parameter values where stability is lost in one or more transversal directions when changing $\epsilon$, $\chi_{dyn}$, or the regular graph's properties, such as its cyclic symmetry, size $N$, or degree $k$.

\begin{figure}[htbp]
    \centering
    \includegraphics[width=0.49\columnwidth]{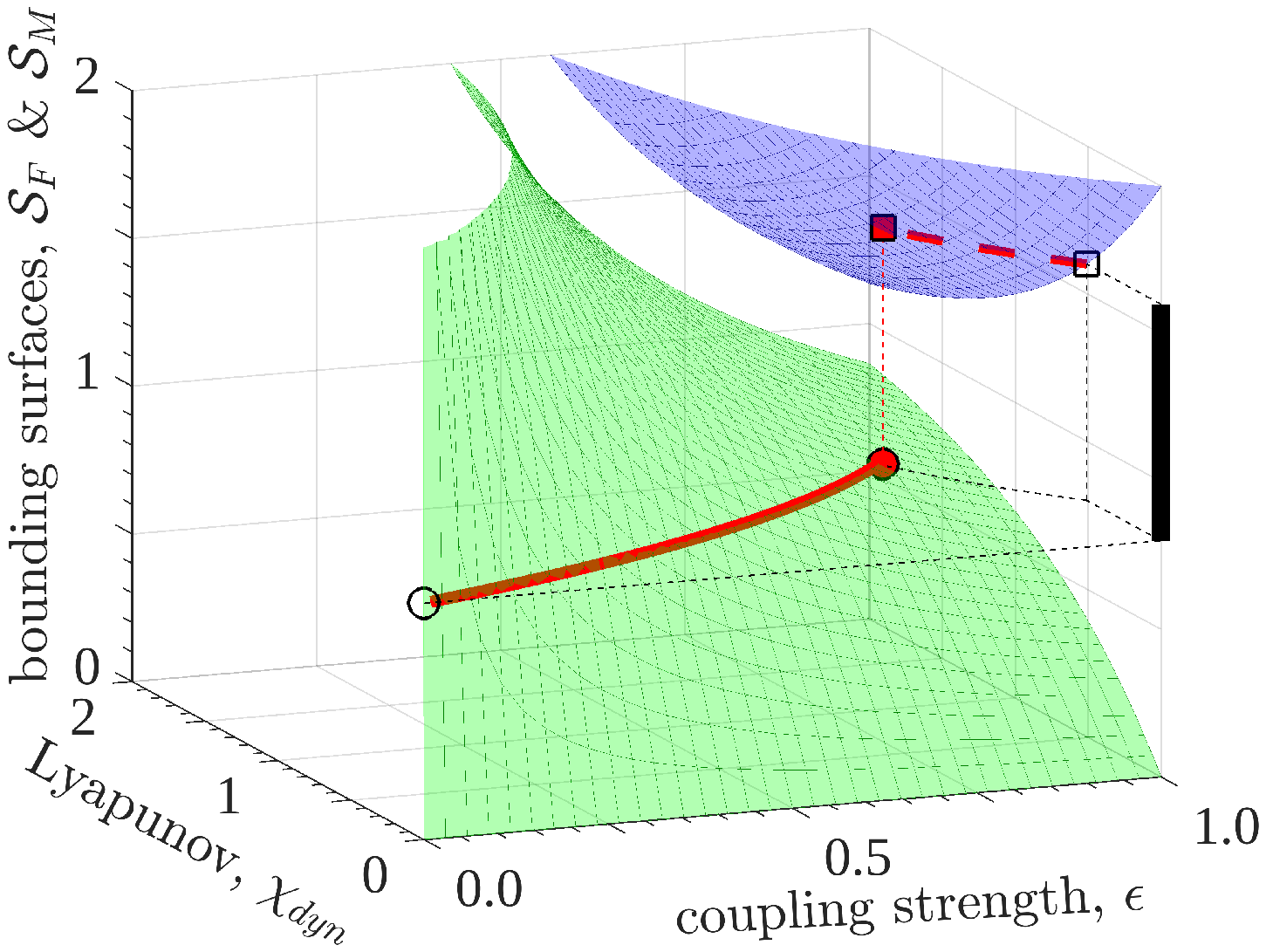}
    \includegraphics[width=0.49\columnwidth]{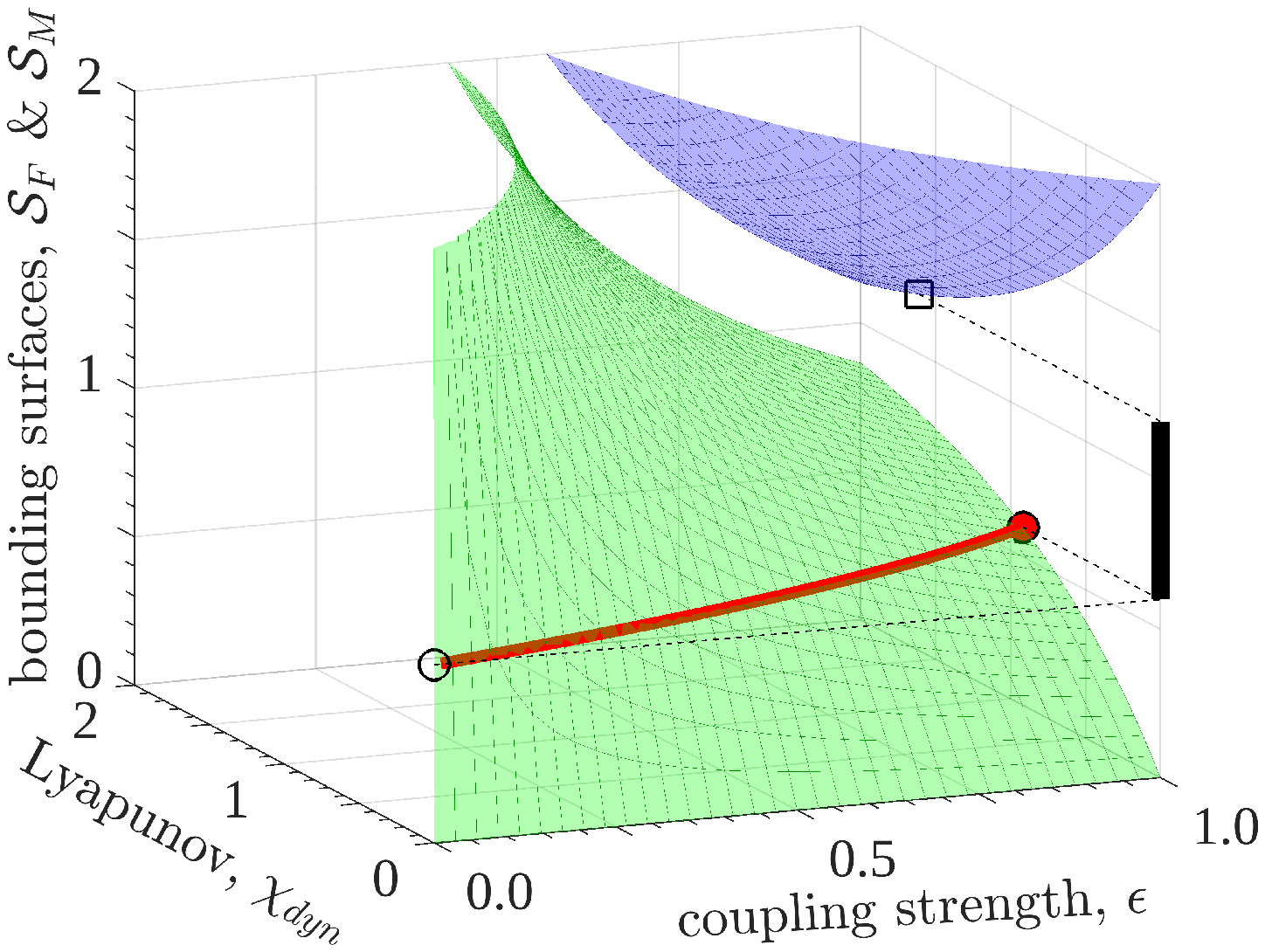}
    \caption{ {\bf Stability criteria for the synchronisation of identical, chaotic maps, coupled diffusively in generic regular graphs}. Linearly stable synchronisation happens as long as the graph's normalised minimum non-zero and maximum Laplacian eigenvalues, $\lambda_F/k$ and $\lambda_M/k$ (vertical interval at the $\epsilon = 1$ plane), fit between $\mathcal{S}_F$ (bottom) and $\mathcal{S}_M$ (top) surfaces. The minimum coupling strength needed to synchronise the maps, $\epsilon^{(c)}$, is defined by the intersection of $\lambda_F/k$ with $\mathcal{S}_F$ and depends on the map's Lyapunov exponent, $\chi_{dyn}$ (continuous curve in both panels). As $\epsilon$ increases, $\lambda_M/k$ can intersect $\mathcal{S}_M$ (dashed curve in left panel), defining a maximum Lyapunov exponent, $\chi_{dyn}^{max}$ (filled symbols), where the synchronisation manifold then looses stability if $\chi_{dyn}$ or $\epsilon$ are increased.}
    \label{fig_StabCriteria}
\end{figure}

The $2$ bounding surfaces in Eq.~(\ref{eq:stability_cond}) -- $\mathcal{S}_F(\chi_{dyn},\,\epsilon)$ and $\mathcal{S}_M(\chi_{dyn},\,\epsilon)$ -- create $2$ scenarios depending on the regular graph's $\lambda_F/k$ and $\lambda_M/k$ possibility to intersect the surfaces as $\epsilon$ or $\chi_{dyn}$ change, 
which we illustrate in Fig.~\ref{fig_StabCriteria}. A critical curve is defined in the lower bounding surface at the height where $\lambda_F/k$ intersects $\mathcal{S}_F(\chi_{dyn},\,\epsilon)$. Similarly, a critical curve for the upper bounding surface is defined at the intersection of $\lambda_M/k$ with $\mathcal{S}_M(\chi_{dyn},\,\epsilon)$.

The case shown on the left panel in Fig.~\ref{fig_StabCriteria} corresponds to regular graphs where the critical curves share a common crossing $(\epsilon^{(c)},\,\chi_{dyn}^{max})$ at $\epsilon^{(c)}\in(0,\,1]$; highlighted by filled symbols in the panel. This crossing happens when $[1 - \exp(\chi_{dyn}^{max})]/(\lambda_F/k) = \epsilon^{(c)} = [1 + \exp(\chi_{dyn}^{max})]/(\lambda_M/k)$, where a $\chi_{dyn} > \chi_{dyn}^{max}$ or $\epsilon > \epsilon^{(c)}$ destabilises synchronisation. This crossing allows us to derive the \emph{maximum chaoticity that can be stably synchronised} in these cyclic graphs,
\begin{equation}
    \chi_{dyn}^{max} \equiv -\log\left[ \frac{1 - (\lambda_F/\lambda_M)}{1 + (\lambda_F/\lambda_M)}\right] = 2\tanh^{-1}\left(\frac{\lambda_F}{\lambda_M}\right).
    \label{eq_CriticLyapM_1}
\end{equation}
It is worth noting that this upper limit for the Lyapunov exponent, $\chi_{dyn}^{max}$, is sometimes missed in synchronisation research.

The case shown on the right panel in Fig.~\ref{fig_StabCriteria} corresponds to regular graphs where the crossing is absent (happens outside the $\epsilon\in[0,\,1]$ range). In this case, as $\epsilon$ is increased from $0$ to $1$ and $\chi_{dyn}$ is increased according to the lower bounding surface critical curve, $[1 - \exp(\chi_{dyn}^{F})]/\epsilon_F = \lambda_F/k$, the upper bounding surface is not crossed by $\lambda_M$. Consequently, the \emph{maximum chaoticity that can be stably synchronised} is
\begin{equation}
 \chi_{dyn}^{max} \equiv \chi_{dyn}^{F} = -\log\left[ 1 - \frac{\lambda_F}{k} \right],
    \label{eq_CriticLyapM_2}
\end{equation}
which is highlighted by a filled circle in the right panel at the $\epsilon = 1$ plane.

We can now define a \emph{set of critical regular graphs dividing these $2$ classes of regular graphs}. We do this by matching Eqs.~(\ref{eq_CriticLyapM_1}) and (\ref{eq_CriticLyapM_2}) to find a relationship between $\lambda_F/k$ and $\lambda_M/k$; that is, $(1 - \lambda_F^{(c)}/\lambda_M^{(c)})/(1 + \lambda_F^{(c)}/\lambda_M^{(c)}) = (1 - \lambda_F^{(c)}/k)$,
\begin{equation}
    \frac{\lambda_F^{(c)}}{k} + \frac{\lambda_M^{(c)}}{k} = 2.
    \label{eq_CriticEigs}
\end{equation}
This general distinction shows that the left panel in Fig.~\ref{fig_StabCriteria} corresponds to regular graphs that hold $\lambda_F/k + \lambda_M/k > 2$ [and stability follows Eq.~(\ref{eq_CriticLyapM_1})] and the right panel in Fig.~\ref{fig_StabCriteria} corresponds to regular graphs that hold $\lambda_F/k + \lambda_F/k < 2$ [and stability follows Eq.~(\ref{eq_CriticLyapM_2})]. The critical set of regular graphs -- those fulfilling Eq.~(\ref{eq_CriticEigs}) -- can be analysed by either Eq.~(\ref{eq_CriticLyapM_1}) or (\ref{eq_CriticLyapM_2}).

We note that for any graph, $\lambda_F/k\in(0,\,N/(N-1)]$ and $\lambda_M/k\in[N/(N-1),\,2]$ \cite{Hahn_1997,Das_2012}. This means that $N/(N-1) < \lambda_F/k + \lambda_M/k \leq N/(N-1) + 2 = (3N-1)/(N-1)$ always. For example, a complete graph, $\mathcal{C}_N(k=N-1)$ (i.e., a cyclic graph with $k = N-1$ defining an all-to-all coupling) has $\lambda_F/k = \lambda_M/k = N/(N-1)$, hence, $\lambda_F/k + \lambda_M/k = 2N/(N-1) > 2$. This means that complete graphs belong to the case from our left panel in Fig.~\ref{fig_StabCriteria}, and according to Eq.~(\ref{eq_CriticLyapM_1}),  $\chi_{dyn}^{max}[\mathcal{C}_N(k=N-1)] = \infty$, which means that they can stably synchronise any chaotic map.

In both classes of regular graphs, the \emph{minimum coupling strengths needed to maintain a linearly stable synchronisation} for different Lyapunov exponents, is given by the critical curve $\mathcal{S}_F(\epsilon^{(c)},\chi_{dyn}) = [1 - \exp(\chi_{dyn})]/\epsilon^{(c)} = \lambda_F/k$, and is valid up to $\chi_{dyn}^{max}$ -- depending on the regular graph, either from Eq.~(\ref{eq_CriticLyapM_1}) or Eq.~(\ref{eq_CriticLyapM_2}). Namely,
\begin{equation}
    \epsilon^{(c)} = \left[ 1 - \exp(-\chi_{dyn}) \right] \left(\frac{\lambda_F}{k}\right)^{-1}\;\forall\,\chi_{dyn}\in(0,\,\chi_{dyn}^{max}].
    \label{eq_CriticCoupling}
\end{equation}
This curve is shown in both panels of Fig.~\ref{fig_StabCriteria} by a thick continuous line.
%
    \subsection*{Synchronisation-Manifold's Stability for Specific Cyclic Graphs}
In what follows, we derive closed-form expressions for the critical points of the synchronisation-manifold's stability [Eqs.~(\ref{eq_CriticLyapM_1})-(\ref{eq_CriticCoupling})] in $2$ specific cyclic graphs, including their critical link densities, $\rho_c = k/(N-1)$. We focus on $k$-cycle graphs, $\mathcal{C}_N(k)$, and $k$-M{\"o}bius ladders, $\mathcal{M}_N(k)$. $\mathcal{C}_N(k)$ are cyclic graphs with even degrees where connections span $k$ neighbours per node in ring-like structure (also known as Wiley-Strogatz-Girvan networks \cite{Strogatz_2006}). $\mathcal{M}_N(k)$ are our generalisation of the M{\"o}bius ladder \cite{mob_lad_org,mob_lad}, which has $k = 3$. We introduce $\mathcal{M}_N(k)$ graphs to increase the degree to $3 \leq k \leq N-1$, but keeping their overall ladder-like topology. Our derivations for $\mathcal{C}_N(k)$ and $\mathcal{M}_N(k)$ include finite size critical points and thermodynamic limits.
%
    \subsubsection*{Results for $k$-cycles.} These graphs have degrees $k = 2q$, with $q\in\mathbb{N}>0$, and can be represented by a Laplacian matrix, $\mathbf{L}[\mathcal{C}_N(k=2q)]$, whose first row is given by
\begin{equation}
    L_{1j} = \left\lbrace \begin{array}{ll}
                  k & \textnormal{if} \;\; j=1, \\
                  -1 & \textnormal{if} \;\; j=2,\dots,\frac{k}{2}+1,\;\;\\
                  -1 & \textnormal{if} \;\; j=N,\dots,N - (\frac{k}{2}-1),\\
                  0 & \textnormal{otherwise}.
                \end{array} \right.
    \label{eq_kcycles_def}
\end{equation}
\begin{figure}[ht]
    \centering
    \includegraphics[width=0.49\columnwidth]{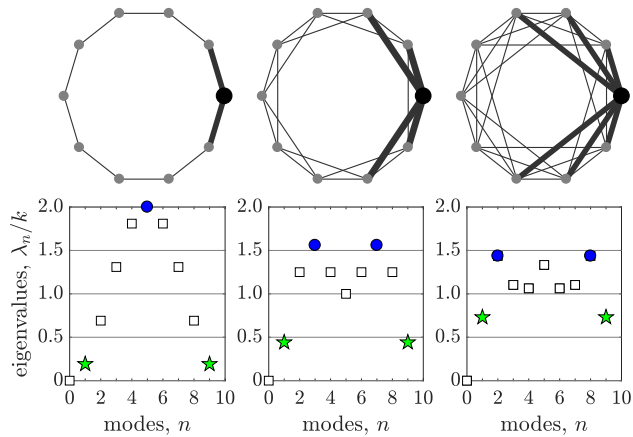}
    \caption{\textbf{$10$-node $k$-cycle graphs with normalised Laplacian eigenvalues}. From left to right, top panels show a $2$-cycle ($k = 2$), a $4$-cycle ($k = 4$), and a $6$-cycle graph ($k = 6$), where a node's neighbourhood is highlighted by thick lines. Bottom panels show the respective normalised Laplacian eigenvalues, where the minimum non-zero (Fiedler) and maximum eigenvalues are highlighted by stars and circles, respectively.}
    \label{fig_k_graphs}
\end{figure}

Because of the cyclic property and the cosine symmetry in Eq.~(\ref{eq:eigenvals}), we find that the eigenvalues for $\mathcal{C}_N(k)$ (see Appendix: \hyperlink{k-cycles}{$k$-cycles}) are given by
\begin{equation}\label{eq:eigenvals_kcycles}
   \lambda_n[\mathcal{C}_N(k)] = k - 2\sum_{s=1}^{k/2}\cos\left(\frac{2\pi n}{N}s\right) = k + 1 - \left[\frac{ \sin\left(\frac{n\,\pi (k+1)}{N}\right) }{ \sin\left(\frac{n\,\pi}{N}\right) }\right]\!.
\end{equation}
Equation~(\ref{eq:eigenvals_kcycles}) is valid for $n = 0,\ldots,N-1$, since it can be shown by trigonometric identities that $\sin\left[\pi n (k+1)/N\right]/\sin\left(\pi n/N\right) = k+1$ when $n = 0$. For example, Fig.~\ref{fig_k_graphs} shows three examples of $k$-cycles and their respective eigenvalue spectra -- from left to right, $\mathcal{C}_{10}(2)$, $\mathcal{C}_{10}(4)$, and $\mathcal{C}_{10}(6)$ -- where we highlight (by stars) that the first non-zero eigenvalues is doubly degenerated.

In order to find the critical points for the local stability of the synchronisation manifold, we need the smallest and largest eigenvalues from Eq.~(\ref{eq:eigenvals_kcycles}), $\lambda_F$ and $\lambda_M$, respectively. For any degree $k = 2q$, we find that (see Eqs.~(\ref{eq_kCycleMinEig}) and (\ref{eq_kCycleMaxEig}) in Appendix: \hyperlink{k-cycles}{$k$-cycles}) these eigenvalues correspond to
\begin{eqnarray}
    \lambda_{F}[\mathcal{C}_N(k)] = \min_{n>0}\{\lambda_n\} = \lambda_1 = k + 1 - \frac{\sin\left( \pi(k+1)/N \right) }{ \sin\left(\pi/N\right)}, 
    \label{eq_k-cycle_lambdaF} \\ \nonumber
    \lambda_{M}[\mathcal{C}_N(k)] = \max_{n>0}\{\lambda_n\} = \max\{\lambda_{\lfloor 3N/2(k+1)\rfloor},\,\lambda_{\lceil 3N/2(k+1) \rceil}\} = k + 1 - \\
    \min\left\lbrace \frac{ \sin\left( \lfloor \frac{3N}{2(k+1)}\rfloor \frac{\pi(k+1)}{N} \right) }{ \sin\left( \lfloor \frac{3N}{2(k+1)}\rfloor \frac{\pi}{N} \right) },\, \frac{ \sin\left( \lceil \frac{3N}{2(k+1)}\rceil \frac{\pi(k+1)}{N} \right) }{ \sin\left( \lceil \frac{3N}{2(k+1)}\rceil \frac{\pi}{N} \right) } \right\rbrace,
    \label{eq_k-cycle_lambdaM}
\end{eqnarray}
where $\lfloor\cdot\rfloor$ rounds the argument down to the next smaller integer and $\lceil\cdot\rceil$ rounds the argument up to the next larger integer.

We note that for large $k$-cycles with non-vanishing link densities, $\rho$, Eq.~(\ref{eq_k-cycle_lambdaF}) can be approximated to $\lambda_F/k \simeq 1 - \textnormal{sinc}\left[ \pi (k+1)/N \right]$, where $\textnormal{sinc(x)} = \sin(x)/x$ and $(k+1)/k\to1$. This implies that in the limit of $N\to\infty$ and $\rho = k/(N-1)$ finite, $\lambda_F/k\to 1 - \textnormal{sinc}(\pi\,\rho) < 1$. On the other hand, Eq.~(\ref{eq_k-cycle_lambdaM}) approximates to $\lambda_{M}/k \simeq 1 - \sin(3\pi/2)/(k+1)\sin[3\pi/2(k+1)]$ for large $k$-cycles with non-vanishing $\rho$, and $\lambda_{M}/k \to 1 + 2/3\pi > 1$ when $N\to\infty$.

More importantly, according to Eqs.~(\ref{eq_k-cycle_lambdaF}) and (\ref{eq_k-cycle_lambdaM}), $k$-cycles with $2 < k < k_{\mathcal{C}}$ are such that $\lambda_M/k + \lambda_F/k < 2$, $k_\mathcal{C}$ being the \emph{critical $k$-cycle degree} that makes $\lambda_M/k_\mathcal{C} + \lambda_F/k_\mathcal{C} = 2$ (see Eq.~(\ref{eq_kc_kcycles}) in Appendix \hyperlink{k-cycles}{$k$-cycles}). This implies that most $k$-cycles belong to the class of cyclic graphs with a $\chi_{dyn}^{max}[\mathcal{C}_N(k)]$ given by Eq.~(\ref{eq_CriticLyapM_2}) -- with the exception of the ring graph, $\mathcal{C}_N(2)$, and the nearly complete $k$-cycles, $\mathcal{C}_N(k\geq k_\mathcal{C})$. Hence, the \emph{maximum chaoticity that can be stably synchronised in a $k$-cycle} with $2 < k < k_{\mathcal{C}}$ is
\begin{equation}
 \chi_{dyn}^{max}[\mathcal{C}_N(k)] = - \log\left[ \frac{ \sin\!\left[\pi(k+1)/N\right] }{ k\sin(\pi/N) } - \frac{1}{k} \right],
    \label{eq_chi_k-cycle}
\end{equation}
which is determined from Eq.~(\ref{eq_CriticLyapM_2}) by substituting $\lambda_F$ from Eq.~(\ref{eq_k-cycle_lambdaF}).

We note that for a fixed size, $N$, the maximum Lyapunov exponent in Eq.~(\ref{eq_chi_k-cycle}) grows as a function of the degrees as power law with exponent $2$, i.e., $\chi_{dyn}^{max}[\mathcal{C}_N(k)] \sim k^2$. In terms of $\rho$, Eq.~(\ref{eq_chi_k-cycle}) holds in the thermodynamic limit ($N\to\infty$ while $\rho$ finite)
\begin{equation}
 \chi_{dyn}^{max}[\mathcal{C}_\infty(\rho)] = - \log\left[ \!\textnormal{ sinc}\!\left(\pi\,\rho\right) \right].
    \label{eq_chi_infinite-cycle}
\end{equation}

The $k$-cycles falling outside this degree range, i.e., with $k = 2$ or $k > k_{\mathcal{C}}$, have a $\chi_{dyn}^{max}[\mathcal{C}_N(k)]$ determined by Eq.~(\ref{eq_CriticLyapM_1}), which requires both $\lambda_F$ and $\lambda_M$ expressions. This set of $k$-cycles becomes vanishingly small on the infinite limit size because $k_\mathcal{C}\to N-1$.

We can now derive an explicit expression for the \emph{minimum coupling strength, $\epsilon^{(c)}[\mathcal{C}_N(k),\,\chi_{dyn}]$, necessary to sustain a locally-stable complete synchronisation in $k$-cycles} by substituting Eq.~(\ref{eq_k-cycle_lambdaF}) into Eq.~(\ref{eq_CriticCoupling}). That is,
\begin{equation}
 \epsilon^{(c)}[\mathcal{C}_N(k),\,\chi_{dyn}] = \frac{k\,\left[1 - \exp(-\chi_{dyn}) \right]}{k + 1 - \sin\!\left(\pi(k+1)/N\right)/\sin(\pi/N)},
    \label{eq_eps_k-cycle}
\end{equation}
which is valid if $\chi_{dyn} < \chi_{dyn}^{max}[\mathcal{C}_N(k)]$. In the thermodynamic limit and if $\chi_{dyn} < \chi_{dyn}^{max}[\mathcal{C}_\infty(\rho)]$, then
\begin{equation}
 \epsilon^{(c)}[\mathcal{C}_\infty(\rho),\,\chi_{dyn}] = \frac{ 1 - \exp(-\chi_{dyn}) }{1 - \textnormal{sinc}\!\left(\pi\,\rho\right)}.
    \label{eq_eps_infinite-cycle}
\end{equation}

\begin{figure}[htbp]
    \centering
    \includegraphics[width=0.49\columnwidth]{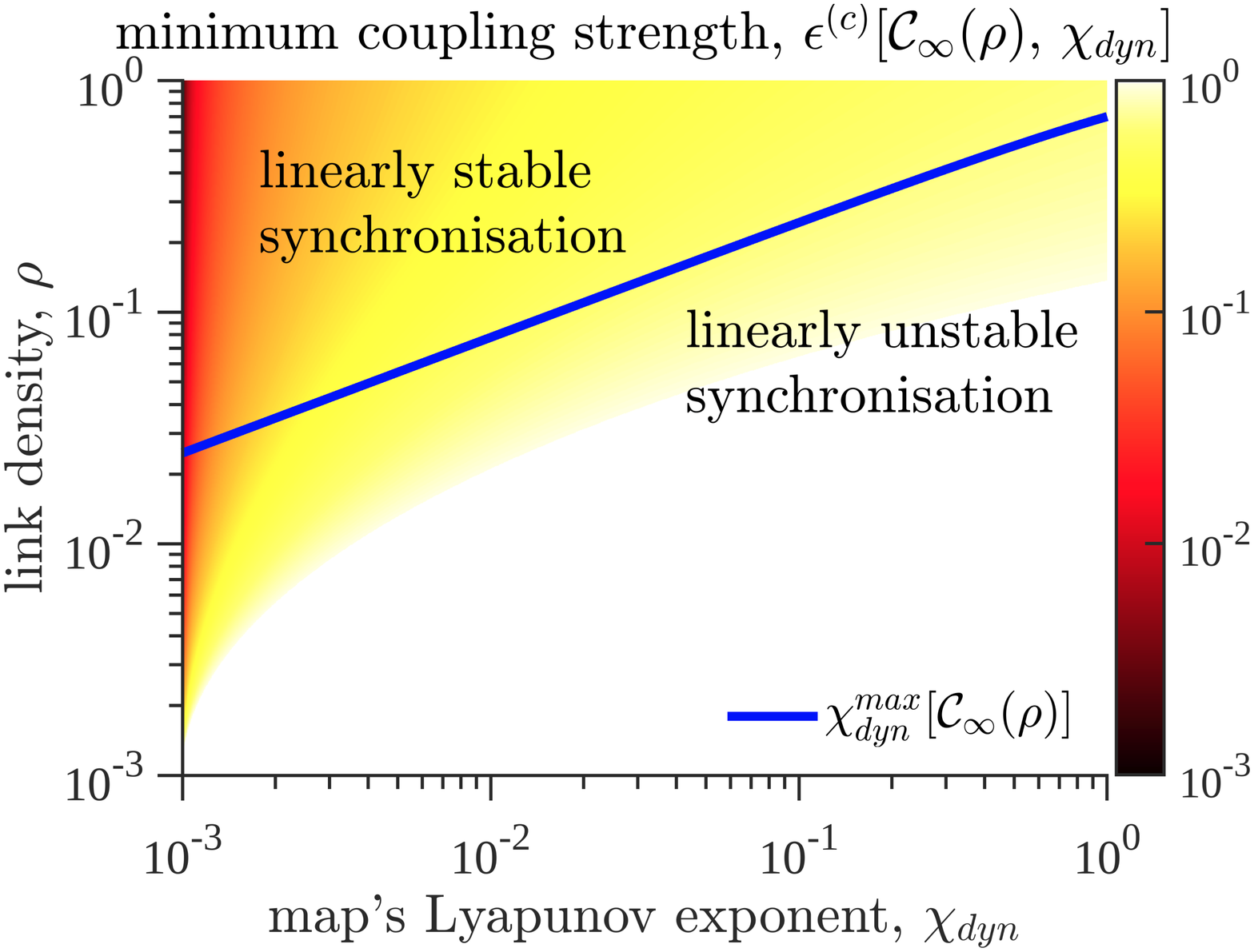}
    \includegraphics[width=0.49\columnwidth]{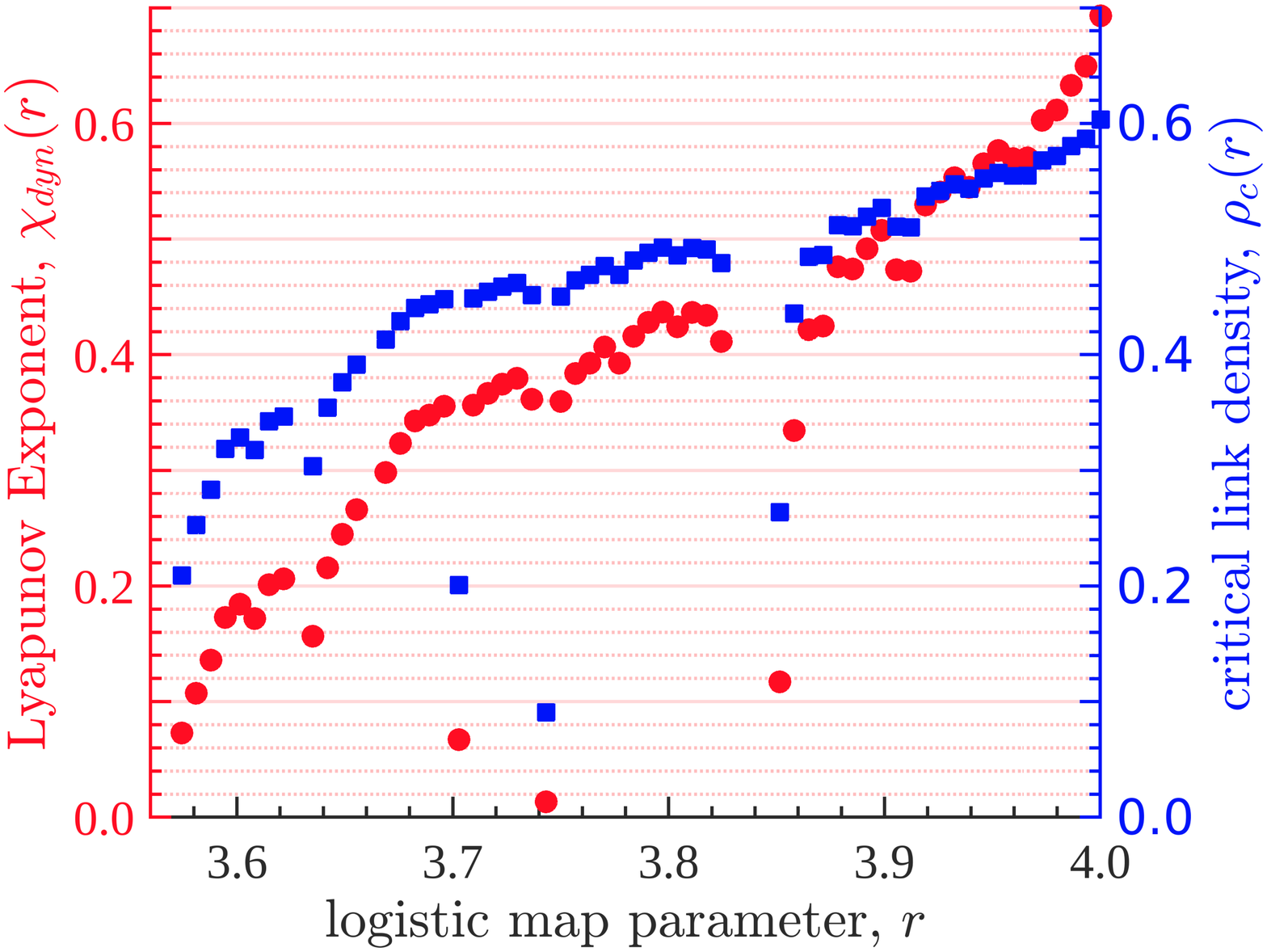}
    \caption{\textbf{Critical stability points of the synchronisation manifold for infinitely large $k$-cycles of identical maps}. Left panel shows in colour code, the minimum coupling strength, $\epsilon^{(c)}$ [Eq.~(\ref{eq_eps_infinite-cycle})], needed to sustain a linearly-stable synchronisation as a function of the link density, $\rho$, and map's Lyapunov exponents, $\chi_{dyn}$. The line signals the maximum chaoticity, $\chi_{dyn}^{max}$ [Eq.~(\ref{eq_chi_infinite-cycle})], that can be stably synchronised in such a $k$-cycle. Right panel shows a numerical example for logistic maps, coupled in $k$-cycles. Filled (red) circles are the isolated map's Lyapunov exponent, $\chi_{dyn}(r)$, as a function of the map's parameter, $r$, and filled (blue) squares show our thermodynamic-limit prediction for the critical (minimum) link-density, $\rho_c$ (non-chaotic solutions, i.e., $\chi_{dyn} \leq 0$, are excluded).}
    \label{fig:k-cycle_CriticPnts}
\end{figure}

We note that Eq.~(\ref{eq_eps_infinite-cycle}) would hold $\epsilon_\infty^{(c)} > 1$ whenever $\textnormal{sinc}(\pi\,\rho) > \exp(-\chi_{dyn})$; but this is an unstable state that happens when the map's Lyapunov exponent is such that $\chi_{dyn} > \chi_{dyn}^{max}[\mathcal{C}_\infty(\rho)]$ for a given infinite-sized $k$-cycle. $\epsilon^{(c)}[\mathcal{C}_\infty(\rho),\,\chi_{dyn}]$ can be seen on the left panel of Fig.~\ref{fig:k-cycle_CriticPnts} in logarithmic scale and in colour code, where $\chi_{dyn}^{max}[\mathcal{C}_\infty(\rho)]$ is signaled by a thick dashed (diagonal) line. Below this line, the synchronisation becomes linearly unstable, which corresponds to $k$-cycles with sparse connections and maps with Lyapunov exponent greater than $\chi_{dyn}^{max}[\mathcal{C}_\infty(\rho)]$.

Using the thermodynamic limit from Eq.~(\ref{eq_chi_infinite-cycle}), we can derive the \emph{minimum link density needed to sustain a linearly-stable synchronisation} in infinite-sized $k$-cycles of chaotic maps, which is given by
\begin{equation}\label{eq:cry_conect}
    \rho_c = \frac{1}{\pi}\textnormal{sinc}^{-1}(\exp(-\chi_{dyn})),\;\;\textnormal{for}\;0 < \chi_{dyn} < \chi_{dyn}^{max}[\mathcal{C}_\infty(\rho)].
\end{equation}
This implies that it is necessary that $\rho\geq\rho_c$ in order to sustain a locally-stable synchronisation for an infinite number of coupled maps with Lyapunov exponent $\chi_{dyn}$. For example, if we take $\chi_{dyn} = \log(2)$ (as in a fully chaotic logistic, tent, or shift map), Eq.~(\ref{eq:cry_conect}) results in $\rho_c = \textnormal{sinc}^{-1}(2)/\pi \simeq 0.60335$, which is a dense $k$-cycle. In practical situations, we can use Eq.~(\ref{eq:cry_conect}) to find $\rho_c$ as a function, for example, of the logistic map's control parameter, $r$, as it is shown on the right panel of Fig.~\ref{fig:k-cycle_CriticPnts}. In this way, we can compare the changes in $\chi_{dyn}(r)$ with the changes in $\rho_c(r)$ as we decrease $r$. As expected, we find that the $k$-cycle can be less densely connected and still maintain a linearly-stable synchronisation manifold, i.e., $\rho_c(r<4) < \rho_c(r=4)$.
%
    \subsubsection*{Results for $k$-M{\"o}bius ladders.} 
\begin{figure}[htbp]
    \centering
    \includegraphics[width=0.49\columnwidth]{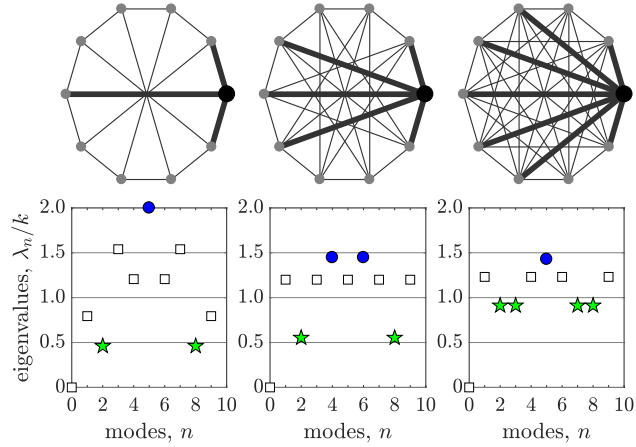}
    \caption{\textbf{$10$-node $k$-M{\"o}bius ladders with normalised Laplacian eigenvalues}. From left to right, the top panels show a M{\"o}bius ladder ($k = 3$), and $2$ generalisations, the $5$-M{\"o}bius ladder ($k = 5$) and the $7$-M{\"o}bius ladder ($k = 7$). Bottom panels show their respective normalised Laplacian eigenvalues (stars signal the Fiedler eigenvalue and circles the maximum eigenvalue) as in Fig.~\ref{fig_k_graphs}.}
    \label{fig_kML_graphs}
\end{figure}

These cyclic graphs are a generalisation of the M{\"o}bius ladder. M{\"o}bius ladders are cyclic graphs with either $k = 3$ or $4$ neighbours \cite{mob_lad_org,mob_lad}, making them equivalent to the M{\"o}bius strip -- a two-dimensional, non-orientable, manifold. A M\"{o}bius ladder with $k =3$ can be constructed, for example, by adding $N/2$ new links (with $N>3$ and even) connecting opposite nodes of a $2$-cycle known as \emph{rungs}; as it can be seen on the left panel in Fig.~\ref{fig_kML_graphs}. However, M{\"o}bius ladders have a vanishing link density, $\rho$, when $N\to\infty$. We introduce here a way to construct $k$-M{\"o}bius ladders, $\mathcal{M}_N(k)$, with arbitrary $k$, keeping $\rho$ finite when $N\to\infty$.

We generalise \emph{rungs} by adding $k - 2$ edges to each node of a $2$-cycle (i.e., a ring, $\mathcal{C}_N(2)$), making these edges connect each node to its $k-2$ furthest nodes in a $2$-cycle. Our construction is restricted to have $N$ odd [even] if $k$ is even [odd], which is fulfilled whenever $N + 5 - k = 2q$, with $q \in\mathbb{N} > 2$ and $k \leq N-1$ (the left panel in Fig.~\ref{fig_kML_graphs} has $N(k=3,\,q=6) = 3 - 5 + 2\times6 = 10$). The first row of $\mathbf{L}[\mathcal{M}_N(k)]$ is then given by
\begin{equation}
    L_{1j} =\left\lbrace \begin{array}{ll}
                  k & \textnormal{if} \;\; j=1, \\
                  -1 & \textnormal{if} \;\; j=2,N-1 \;\; \textnormal{($2$-cycle edges)}, \\
                  -1 & \textnormal{if} \;\; j=(N+5-k)/2,\ldots,(N-1+k)/2, \\
                  0 & \textnormal{otherwise}.
                \end{array} \right.
    \label{eq_kMobius_def}
\end{equation}

We find a compact expression for the Laplacian eigenvalues for $k$-M{\"o}bius ladders by substituting Eq.~(\ref{eq_kMobius_def}) in Eq.~(\ref{eq:eigenvals}) (see Appendix: \hyperlink{k-ML}{$k$-M{\"o}bius ladders}),
\begin{eqnarray}\nonumber
      \lambda_n[\mathcal{M}_N(k)] = k - 2\cos\left(\frac{2\pi n}{N}\right) - \sum_{j=(N-k+5)/2}^{(N-1+k)/2}\cos\left(\frac{2\pi n(j-1)}{N}\right) = \\
      k + 1 - \left[ \frac{ \sin\left(3\pi\,n/N\right) + (-1)^n \sin\left(\pi\,n(k-2)/N\right) }{\sin\left(n\,\pi/N\right)} \right].
    \label{eq:eigenval_gen_mob_lad}
\end{eqnarray}
From Eq.~(\ref{eq:eigenval_gen_mob_lad}), it can be shown that $\lambda_{M}[\mathcal{M}_N(k)] = \max_{n}\{\lambda_n\} = \lambda_1$ if $7 \leq k \leq N-1$, and that $\lambda_{F}[\mathcal{M}_N(k)] = \min_{n}\{\lambda_n>0\} = \lambda_2$ if $3 \leq k \leq k_c \simeq (2N+8)/5$ (see Appendix: \hyperlink{k-ML}{$k$-M{\"o}bius ladders}). Outside these ranges, $\lambda_F$ and $\lambda_M$ change to other modes. Focusing on these ranges, when $\lambda_M[\mathcal{M}_N(k)] = \lambda_1$ we have
\begin{equation}
    \lambda_{M}[\mathcal{M}_N(k)] = \lambda_1 = k + 1 - \left[ \frac{  \sin\left(3\pi/N\right) - \sin\left(\pi\,(k-2)/N \right) }{\sin\left(\pi/N\right)} \right]\!,
    \label{eq_k-ML_lambdaM}
\end{equation}
which for $N\to\infty$ and $\rho$ non-diluted (i.e., avoiding small $\rho$ such that $k\geq7$)
\begin{equation}
    \frac{\lambda_1[\mathcal{M}_N(k)]}{k} \simeq 1 - \frac{3}{2\rho(N-1)} + \frac{\sin(\pi\rho)}{\pi\rho} \to 1 + \textnormal{sinc}(\pi\rho).
    \label{eq:lamda_M_mob}
\end{equation}

On the other hand, when $3 \leq k \leq k_c \simeq (2N+8)/5$ and $\lambda_F[\mathcal{M}_N(k)] = \lambda_2$,
\begin{equation}
    \lambda_{F}[\mathcal{M}_N(k)] = k + 1 - \left[ \frac{  \sin\left(6\pi/N\right) + \sin\left(2\pi\,(k-2)/N \right) }{\sin\left(2\pi/N\right)} \right]\!,
    \label{eq_k-ML_lambdaF}
\end{equation}
which for $N\to\infty$ and $\rho < \rho_c \simeq 2/5$ (i.e., diluted or avoiding large $\rho$)
\begin{equation}
    \frac{\lambda_2[\mathcal{M}_N(k)]}{k} \simeq 1 - \frac{3}{\rho(N-1)} - \frac{\sin(2\pi\rho)}{2\pi\rho} \to 1 - \textnormal{sinc}(2\pi\rho).
    \label{eq:lamda_F_mob}
\end{equation}

\begin{figure}[htbp]
    \centering
    \includegraphics[width=.49\columnwidth]{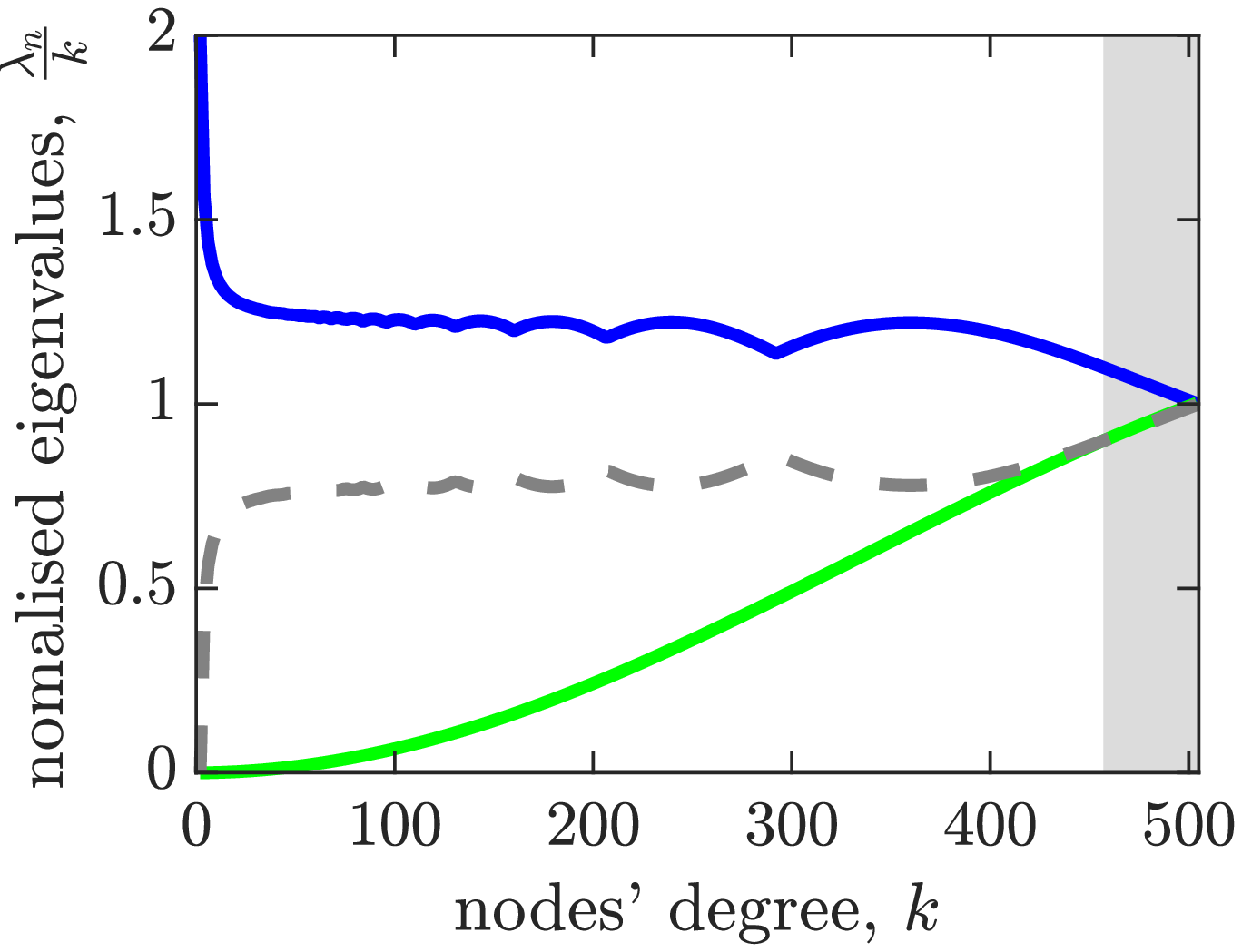}
    \includegraphics[width=.49\columnwidth]{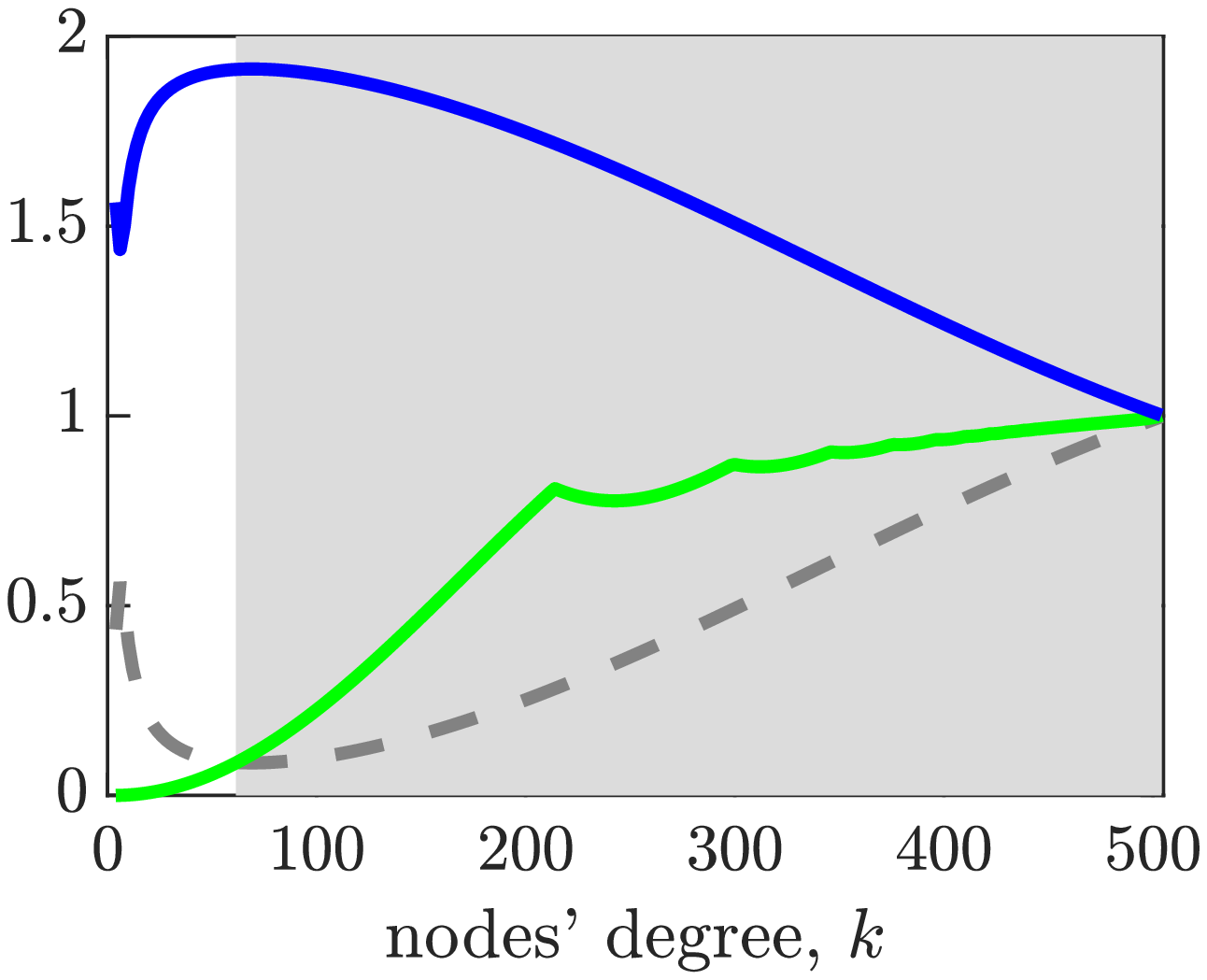}
    \caption{\textbf{Maximum and minimum normalised eigenvalues for $k$-cylces (left) and $k$-M{\"o}bius ladders (right) with $505$ nodes.} The blue [green] line corresponds to the maximum [minimum] normalised eigenvalue $\lambda_M/k$ [$\lambda_F/k$]. Grey dashed lines show $2 - \lambda_M/k$. As long as this distance (dashed lines) is larger than $\lambda_F/k$, $\lambda_M$ plays no role in the synchronisation manifold's stability. However, this distance becomes smaller than $\lambda_F/k$ in the shaded areas on both panels, where the stability is then determined by $\lambda_M/k$.}
    \label{fig:max_and_min_eigenvals}
\end{figure}

According to Eqs.~(\ref{eq_k-ML_lambdaM}) and (\ref{eq_k-ML_lambdaF}), $k$-M{\"o}bius ladders are such that $\lambda_M/k + \lambda_F/k < 2$ (as in the $k$-cycles) when $7\leq k < k_\mathcal{M}$, or $\lambda_M/k + \lambda_F/k > 2$ when $k_\mathcal{M} < k \leq k_c \simeq (2N+8)/5$; $k_\mathcal{M}$ being the \emph{critical M{\"o}bius ladder degree} that makes $\lambda_M/k_\mathcal{M} + \lambda_F/k_\mathcal{M} = \lambda_1/k_\mathcal{M} + \lambda_2/k_\mathcal{M} = 2$. Specifically, $k_\mathcal{M}$ is determined from (see Eq.~(\ref{eq_CriticDegree}) in Appendix: \hyperlink{k-ML}{$k$-M{\"o}bius ladders})
$$ \alpha_N = \frac{\sin\left(2\pi\,(k_\mathcal{M}-2)/N\right)}{\sin(2\pi/N)} - \frac{\sin\left(\pi\,(k_\mathcal{M}-2)/N\right)}{\sin(\pi/N)}, $$
where $\alpha_N = 2 - \sin\left(3\pi/N\right)/\sin(\pi/N) - \sin\left(6\pi/N\right)/\sin(2\pi/N)$. For example, when $N = 505$, as in Fig.~\ref{fig:max_and_min_eigenvals}, we obtain (numerically) that $k_\mathcal{M} \simeq 62$. The maximum and Fiedler eigenvalues for $k$-M{\"o}bius ladders with $\lambda_M/k + \lambda_F/k > 2$ are contained within the shaded area in the right panel of Fig.~\ref{fig:max_and_min_eigenvals}. The remaining cases (in both panels) show the eigenvalues when $\lambda_M/k + \lambda_F/k < 2$.

Consequently, the critical points of the synchronisation manifold's stability in $k$-M{\"o}bius ladders -- maximum Lyapunov exponent that can be synchronised, $\chi_{dyn}^{max}[\mathcal{M}_N(k)]$, and minimum coupling strength, $\epsilon^{(c)}[\mathcal{M}_N(k)]$ -- depend on the degree being smaller or bigger than $k_\mathcal{M}$.
For $7\leq k < k_\mathcal{M}$, the \emph{maximum chaoticity that can be synchronised in $k$-M{\"o}bius ladders} is determined by substituting $\lambda_F[\mathcal{M}_N(k)]$ from Eq.~(\ref{eq_k-ML_lambdaF}) in Eq.~(\ref{eq_CriticLyapM_2}). That is,
\begin{equation}
    \chi_{dyn}^{max}[\mathcal{M}_N(k)] = -\log\left( \frac{  \sin\left(6\pi/N\right) + \sin\left(2\pi\,(k-2)/N \right) }{k\sin\left(2\pi/N\right)} - \frac{1}{k} \right)\!.
    \label{eq_chi2_kML}
\end{equation}
For large $N$, $\lambda_2[\mathcal{M}_\infty(\rho)]/k \simeq 1 - \textnormal{sinc}(2\pi\rho)$ [Eq.~(\ref{eq:lamda_F_mob})]. Hence, the maximum Lyapunov exponent that can be synchronised transforms to
\begin{equation}
    \chi_{dyn}^{max}[\mathcal{M}_\infty(\rho)] \simeq - \log\left[ \!\textnormal{ sinc}\!\left(2\pi\,\rho\right) \right],
    \label{eq_chi_infinite-kML}
\end{equation}
which is valid if $0 < \rho \lesssim k_\mathcal{M}/(N-1)$. However, we note that $k_\mathcal{M}\to0$, meaning that $\chi_{dyn}^{max}$ is only valid for finite-sized $k$-M{\"o}bius ladders. We also note that this expression for $\chi_{dyn}^{max}$ is different from the expression for infinite $k$-cycles [Eq.~(\ref{eq_chi_infinite-cycle})] solely because of the $2$ in the argument of the sinc function.%

For $k_\mathcal{M} < k \leq k_c \simeq (2N+8)/5$, $\chi_{dyn}^{max}[\mathcal{M}_N(k)]$ is determined by substituting $\lambda_F[\mathcal{M}_N(k)]$ and $\lambda_M[\mathcal{M}_N(k)]$ from Eqs.~(\ref{eq_k-ML_lambdaM}) and (\ref{eq_k-ML_lambdaF}) in Eq.~(\ref{eq_CriticLyapM_1}),
\begin{equation}
    \chi_{dyn}^{max}[\mathcal{M}_N(k)] = -\log\left[ \frac{ \lambda_1[\mathcal{M}_N(k)] - \lambda_2[\mathcal{M}_N(k)] }{ \lambda_1[\mathcal{M}_N(k)] + \lambda_2[\mathcal{M}_N(k)] }\right].
    \label{eq_chi1_kML}
\end{equation}
Similarly to $k$-cycles, in the thermodynamic limit ($N\to\infty$) we can define a \emph{critical link density}, $\rho_\mathcal{M}$, for infinite-sized $k$-M{\"o}bius ladders such that $\lambda_1[\mathcal{M}_\infty(\rho)]/k + \lambda_2[\mathcal{M}_\infty(\rho)]/k = 2$, finding that $\rho_\mathcal{M}=0$ or $1$ (see Appendix: \hyperlink{k-ML}{$k$-M{\"o}bius ladders}), meaning that Eq.~(\ref{eq_chi1_kML}) is valid in the range of $k\in(6,N-1]$. Furthermore, $\lambda_1[\mathcal{M}_\infty(\rho)]/k = 1 + \textnormal{sinc}(\pi\rho)$ [Eq.~(\ref{eq:lamda_M_mob})] and $\lambda_2[\mathcal{M}_\infty(\rho)]/k = 1 - \textnormal{sinc}(2\pi\rho)$ [Eq.~(\ref{eq:lamda_F_mob})], meaning that
\begin{equation}
    \chi_{dyn}^{max}[\mathcal{M}_\infty(\rho)] = -\log\left[ \frac{ \textnormal{sinc}(\pi\rho) + \textnormal{sinc}(2\pi\rho) }{ 2 + \textnormal{sinc}(\pi\rho) - \textnormal{sinc}(2\pi\rho) }\right].
    \label{eq_chi_infinite-kML2}
\end{equation}

We can now derive a closed-form expression for the \emph{critical coupling strength necessary to sustain a locally-stable complete-synchronisation in $k$-M{\"o}bius ladders} with $7\leq k \leq k_c \simeq (2N+8)/5$ by substituting $\lambda_2[\mathcal{M}_N(k)]$ into Eq.~(\ref{eq_CriticCoupling}). This results in
\begin{equation}
 \epsilon^{(c)}[\mathcal{M}_N(k),\,\chi_{dyn}] = \frac{ k\,\left[1 - \exp(-\chi_{dyn}) \right] }{ (k + 1) - \left[\frac{ \sin\left(6\pi/N\right) + \sin\left(2\pi\,(k - 2)/N\right) }{\sin(2\pi/N)} \right] },
    \label{eq_eps_k-ML}
\end{equation}
which is valid for $\chi_{dyn}\in(0,\chi_{dyn}^{max}]$, where $\chi_{dyn}^{max}[\mathcal{M}_N(k)]$ is determined from Eq.~(\ref{eq_chi2_kML}) when $7\leq k < k_\mathcal{M}$ and is determined from Eq.~(\ref{eq_chi1_kML}) when $k_\mathcal{M} < k \leq k_c \simeq (2N+8)/5$. In the thermodynamic limit and if $chi_{dyn} < \chi_{dyn}^{max}[\mathcal{M}_\infty(\rho)]$, Eq.~(\ref{eq_eps_k-ML}) transforms to
\begin{equation}
 \epsilon^{(c)}[\mathcal{M}_\infty(\rho),\,\chi_{dyn}] = \frac{ 1 - \exp(-\chi_{dyn}) }{1 - \textnormal{sinc}\!\left(2\pi\,\rho\right)},
    \label{eq_eps_infinite-kML}
\end{equation}
which is similar to the expression for the infinite-sized $k$-cycles from Eq.~(\ref{eq_eps_infinite-cycle}).
%
\section*{Conclusions} 
    \hypertarget{sec_Discuss}{}
In this work, we derive closed-form expressions for the parameters controlling the stability of the synchronisation manifold of identical maps, diffusively coupled in regular graphs -- graphs were all the nodes have the same degree -- and cyclic graphs -- regular graphs with cyclical permutation symmetries. Our detailed derivations are based on the Master Stability Function (MSF) \cite{MSF_pecora,Fujisaka_MSF} and the spectral properties of the graph's Laplacian matrix \cite{CGraphs,GraphSpectra,FanChung} (giving expressions for its eigenvalues), complementing the broad literature of synchronisation in coupled map lattices \cite{Kaneko_1989,Amritkar_1993,Gallas_2004,Lakshmanan_2005,Strogatz_1998,Sync_SW_barahona_pecora,Strogatz_2006} with specific parameter expressions that can be applied straightforwardly.

From the MSF, we study the conditions needed to sustain a stable synchronisation manifold, which require having negative transversal exponents [Eq.~(\ref{eq_StabilityCond})]. We show that these \emph{stability conditions classify regular graphs into two sets} [Fig.~\ref{fig_StabCriteria}]: those that fulfill $\lambda_M/k + \lambda_F/k > 2$ or those that fulfill $\lambda_M/k + \lambda_F/k < 2$, where $\lambda_M$ is the maximum Laplacian eigenvalue, $\lambda_F$ is the minimum non-zero eigenvalue (also known as algebraic connectivity or Fiedler eigenvalue), and $k$ is the graph's degree. The critical set of graphs separating these two sets fulfill $\lambda_M/k + \lambda_F/k = 2$ [Eq.~(\ref{eq_CriticEigs})].

Because of this classification and the MSF conditions, we define critical parameter values. These are the \emph{maximum Lyapunov exponent of the maps}, $\chi_{dyn}^{max}$ [Eqs.~(\ref{eq_CriticLyapM_1}) and (\ref{eq_CriticLyapM_2})] \emph{that can be synchronised holding a linearly stable manifold}, and the \emph{minimum coupling strength}, $\epsilon^{(c)}$ [Eq.~(\ref{eq_CriticCoupling})] \emph{required in generic regular graphs of coupled chaotic maps to synchronise}. Specifically, when $\lambda_F/k + \lambda_M/k < 2$ (as in the non-shaded areas of Fig.~\ref{fig:max_and_min_eigenvals}), the synchronisation's stability and these critical parameters depend solely on $\lambda_F/k$. On the other hand, when $\lambda_F/k + \lambda_M/k > 2$ (as in the shaded areas of Fig.~\ref{fig:max_and_min_eigenvals}) the stability and critical parameters depend on both, $\lambda_F/k$ and $\lambda_M/k$.

We then derive closed-form expressions for the eigenvalues of two specific classes of cyclic graphs: $k$-cycles (i.e., regular lattices with even degree and cyclic symmetry) [Eq.~(\ref{eq:eigenvals_kcycles})] and $k$-M{\"o}bius ladders [Eq.~(\ref{eq:eigenval_gen_mob_lad})], which we introduce to extend the classic M{\"o}bius ladder (which has $k = 3$). From the eigenvalue expressions, we find that $\lambda_F = \lambda_1$ and $\lambda_M = \max\left\lbrace \lambda_{\lfloor 3N/2(k+1) \rfloor}, \lambda_{\lceil 3N/2(k+1) \rceil}\right\rbrace$ for any finite-sized $k$-cycle [Eqs.~(\ref{eq_k-cycle_lambdaF}) and (\ref{eq_k-cycle_lambdaM}), respectively]. However, in $k$-M{\"o}bius ladders, we find that $\lambda_F = \lambda_2$ if $k\in[3,\,k_c]$ (changing to greater modes as $k$ is increased beyond $k_c\simeq(2N+8)/5$) and $\lambda_M = \lambda_1$ if $k\in(6,\,N-1)$ [Eqs.~(\ref{eq_k-ML_lambdaF}) and (\ref{eq_k-ML_lambdaM}), respectively]. From these results, we show that \emph{when the link density is small, both topologies fall into the class of regular graphs where $\lambda_M/k + \lambda_F/k < 2$, but as their density increases, they belong to the other class of regular graphs, where $\lambda_M/k + \lambda_F/k > 2$}. The limits between the sparse and dense regimes, $k_\mathcal{C}$ and $k_\mathcal{M}$ (for $k$-cycles and $k$-M\"{o}bius ladders, respectively), are numerically derived from transcendental equations [Eqs.~(\ref{eq_kc_kcycles}) and (\ref{eq_CriticDegree})]. We also show that for infinite-sized graphs the dependence on the network's degree to determine the stability class disappears.


Having $\lambda_F/k$ and $\lambda_M/k$ in $k$-cycles and $k$-M{\"o}bius ladders, we derive explicit expressions for their critical parameter values in the finite-size and infinite-size limit. Specifically, we determine $\chi_{dyn}^{max}$ for $k$-cycles [Eqs.~(\ref{eq_chi_k-cycle}) and Eq.~(\ref{eq_chi_infinite-cycle}), respectively] and $\epsilon^{(c)}$, as a function of the $k$-cycle properties (i.e., $k$ and $N$ for finite sizes and $\rho$ for infinite sizes) and Lyapunov exponent, $\chi_{dyn}$ [Eqs.~(\ref{eq_eps_k-cycle}) and (\ref{eq_eps_infinite-cycle}), respectively]. Also, we show that these two parameters determine a minimum link density for the synchronisation stability in $k$-cycle [Fig.~\ref{fig:k-cycle_CriticPnts} and Eq.~(\ref{eq_InfCriticDensit})]. Analogously, we carry derivations for $k$-M{\"o}bius ladders [Eqs.~(\ref{eq_chi2_kML}), (\ref{eq_chi_infinite-kML}), (\ref{eq_chi1_kML}), and (\ref{eq_chi_infinite-kML2}) for $\chi_{dyn}^{max}$ and Eqs.~(\ref{eq_eps_k-ML}) and (\ref{eq_eps_infinite-kML}) for $\epsilon^{(c)}$].

We note that other works have derived different properties of the synchronisation manifold of coupled maps and analysed networks with heterogeneity in the node degrees. For example, it has been derived that piece-wise linear, chaotic maps, coupled in a ring lattice, increase their entropy for strong coupling \cite{Viana_2002} (having an overall well-like shape for the coupling strength range). If random connections are added to the lattice (i.e., a Watts-Strogatz model \cite{Watts_1998,Newman_1999,Watts_1999,Newman_2000}), then, the chaoticity of the system decreases with increasing number of random connections \cite{Viana_2007}, which corresponds to the emergence of synchronisation. Instead, if long-range interactions are added, then, the necessary critical coupling-strength for a stable synchronous manifold is known \cite{Viana_2003} (even for non-linear maps), as well as its transient times \cite{Batista_2005}. Similar stability analyses have also been carried out by previous works, describing critical conditions that allow synchronization in coupled map networks \cite{Sync_SW_barahona_pecora,Jost_2001}.

Overall, our work is restricted to regular graphs, which means homogeneous degrees. Because of this restriction, we are able to obtain closed-form expressions for the relevant parameters of the synchronisation-manifold's stability. In spite of this limitation, our results can help in deriving closed-form expressions for other graphs by means of perturbation theory, which would allow to include degree heterogeneity. For example, our $k$-cycle derivations can help when doing perturbation theory on small-world graphs \cite{Watts_1998,Newman_1999,Watts_1999,Newman_2000}.
%
\section*{Appendix}
%
    \subsection*{$k$-cycles -- Minimum and Maximum Laplacian Eigenvalues} \hypertarget{k-cycles}{}
These graphs, $\mathcal{C}_N(k)$, only allow connections between $k$ of the closest neighbours to each node, where $k$ must be an even number. Thus, we write the Laplacian eigenvalues, $\lambda_n[\mathcal{C}_N(k)]$ ($n = 0,\ldots,N-1$), from Eq.~(\ref{eq:eigenvals}) as
\begin{equation}\label{eq:eigenvals_kcycles_sup}
    \lambda_n[\mathcal{C}_N(k)] = \sum_{j=1}^{N} L_{1,j} \cos\!\left(\frac{2 \pi n}{N}(j-1)\right) = k - 2\sum_{s=1}^{k/2}\cos\left(\frac{2\pi n}{N}s\right)\!.
\end{equation}
Here we derive an closed-form expression for the sum on the right-hand-side ($r.h.s.$) of Eq.~(\ref{eq:eigenvals_kcycles_sup}) by expressing the cosine using its complex exponential form. That is
$$ r.h.s. = 2\sum_{s=1}^{k/2}\cos\left(\frac{2\pi n}{N}s\right) = \sum_{s=0}^{k/2}\exp\left[i\frac{2\pi n}{N}s\right] + \sum_{s=0}^{k/2}\exp\left[-i\frac{2\pi n}{N}s\right] - 2, $$
where we replace the $2$ geometric sums by their corresponding results. Namely,
$$ r.h.s. = \frac{ 1 - \exp\left[i\,2\pi\,n (1 + k/2)/N\right] }{ 1 - \exp\left[i\,2\pi\,n/N\right] } + \frac{ 1 - \exp\left[-i\,2\pi\,n(1 + k/2)/N\right] }{ 1 - \exp\left[-i\,2\pi\,n/N\right] } - 2 = \;  $$
$$ \;\; = \frac{ 1 - \exp\left[i\,\pi\,n (k + 2)/N\right] }{ 1 - \exp\left[i\,2\pi\,n/N\right] } + \frac{ 1 - \exp\left[-i\,\pi\,n(k + 2)/N\right] }{ 1 - \exp\left[-i\,2\pi\,n/N\right] } - 2,  $$
which we can transform using the fact that $1 - \exp[\pm\,i\,\phi] = \pm2i\,\sin(\phi/2)\exp[\pm\,i\,\phi/2]$ for an arbitrary phase variable $\phi$. As a result,
$$ r.h.s. = \frac{ 2i\,\sin\left(\pi\,n (k + 2)/2N\right)\, \exp\left[i\,\pi\,n (k + 2)/2N\right] }{ 2i\,\sin\left(\pi\,n/N\right)\,\exp\left[i\,\pi\,n/N\right] } + \;\;$$
$$ + \frac{ (-2i)\,\sin\left(\pi\,n(k + 2)/2N\right)\,\exp\left[-i\,\pi\,n(k + 2)/2N\right] }{ (-2i)\,\sin\left(\pi\,n/N\right)\,\exp\left[-i\,\pi\,n/N\right] } - 2 =\; $$
$$ \;= \frac{\sin\left(\pi\,n(k+2)/2N\right) }{\sin\left(\pi\,n/N\right)}\exp\left[i\,\frac{\pi\,n}{2N}k\right] + \frac{\sin\left(\pi\,n(k+2)/2N\right) }{\sin\left(\pi\,n/N\right)}\exp\left[-i\,\frac{\pi\,n}{2N}k\right]- 2\;\Rightarrow$$
$$ \Rightarrow\;\; r.h.s. = 2\cos\left(\frac{\pi\,n}{2N}k\right)\frac{\sin\left(\pi\,n(k+2)/2N\right) }{\sin\left(\pi\,n/N\right)} - 2. $$
Now, using that $2\cos(\beta)\sin(\alpha) = \sin(\alpha+\beta) + \sin(\alpha-\beta)$ on the $r.h.s.$,
$$ r.h.s. = \frac{ \sin\left(\pi\,n(2k+2)/2N\right) + \sin\left(\frac{\pi\,n}{N}\right)}{ \sin\left(\pi\,n/N\right) } - 2 = \frac{\sin\left(\pi\,n(k+1)/N\right) }{\sin\left(\pi\,n/N\right)} - 1.$$
Finally, our explicit expression for Eq.~(\ref{eq:eigenvals_kcycles_sup}) is
\begin{equation}
    \lambda_n[\mathcal{C}_N(k)] = k - r.h.s. = k + 1 - \frac{ \sin\left(\pi\,n(k+1)/N\right) }{ \sin\left(\pi\,n/N\right) }.
    \label{eq_kcycle_SIexp}
\end{equation}

We note that $\lambda_0[\mathcal{C}_N(k)] = 0$ for any $\mathcal{C}_N(k)$, which can be verified by Eq.~(\ref{eq:eigenvals_kcycles_sup}), and that $\lambda_n[\mathcal{C}_N(k=N-1)] = N,\;\forall\,n>0$ for a complete graph. Also, we observe that $\frac{n}{N}\in[0,\,1),\;\forall\,n$, but because of the cosine in Eq.~(\ref{eq:eigenvals_kcycles_sup}), only the first (non-zero) modes $\frac{n}{N}\in(0,\,1/2]$ are relevant; the remaining $n$ contribute to the eigenvalue degeneracy. As $n$ increases from $1$ up to $\left\lfloor N/2 \right\rfloor$, the denominator in Eq.~(\ref{eq_kcycle_SIexp}) decreases monotonously (without sign changes), making the fraction increasingly larger. Consequently, the smallest non-zero eigenvalue, $\lambda_F$, of any $\mathcal{C}_N(k)$ is its first eigenmode; that is,
\begin{equation}
    \lambda_F[\mathcal{C}_N(k)] \equiv \min_{n\in[1,N/2]}\{\lambda_n[\mathcal{C}_N(k)]\} = \lambda_1.
    \label{eq_kCycleMinEig}
\end{equation}
On the other hand, in order to maximise Eq.~(\ref{eq_kcycle_SIexp}) and find the largest eigenvalue, $\lambda_M$, we can restrict the modes to those that make $\sin\left(n\,\pi(k+1)/N\right)=-1$. The first possible solution is when $n\,\pi(k+1)/N = 3\pi/2$, which is fulfilled when $n = \lfloor 3N/2(k+1) \rceil$ (rounding the argument $3N/2(k+1)$ to the nearest integer). Since $k\in \left[2,N-1\right]$ for any $k$-cycle (implying that $(k+1)\in \left[3,N\right]$), this is a valid solution for the largest Laplacian eigenvalue, $\lambda_M$. Specifically,
\begin{equation}
    \lambda_M[\mathcal{C}_N(k)] = \max\left\lbrace\lambda_{\lfloor 3N/2(k+1) \rfloor},\lambda_{\lceil 3N/2(k+1) \rceil}\right\rbrace.
    \label{eq_kCycleMaxEig}
\end{equation}

Adding the normalised eigenvalues from Eqs.~(\ref{eq_kCycleMinEig}) and (\ref{eq_kCycleMaxEig}), we get
\begin{equation}
    \frac{\lambda_F}{k} + \frac{\lambda_M}{k} = \frac{\lambda_1}{k} + \frac{ \max\left\lbrace\lambda_{\lfloor 3N/2(k+1) \rfloor},\lambda_{\lceil 3N/2(k+1) \rceil}\right\rbrace }{k} ,
    \label{eq_NormEigsSum_kCycle}
\end{equation}
where $ \lambda_F/k + \lambda_M/k<2$ for $2 < k < k_\mathcal{C}$, and $ \lambda_F/k + \lambda_M/k>2$ for $k>k_\mathcal{C}$ (or $k = 2$), being $k_\mathcal{C}$ the critical degree determined by the case when $\lambda_F/k_\mathcal{C} + \lambda_M/k_\mathcal{C}=2$, which explicitly corresponds to
$$  \frac{2}{k_\mathcal{C} }- \frac{1}{k_\mathcal{C} }\left[ \frac{ \sin\left(\pi(k_\mathcal{C}+1)/N\right) }{ \sin\left(\pi/N\right) } \right]$$
$$  - \frac{1}{k_\mathcal{C} } \min\left\lbrace \frac{ \sin\left( \lfloor \frac{3N}{2(k_\mathcal{C}+1)}\rfloor \frac{\pi(k_\mathcal{C}+1)}{N} \right) }{ \sin\left( \lfloor \frac{3N}{2(k_\mathcal{C}+1)}\rfloor \frac{\pi}{N} \right) },\, \frac{ \sin\left( \lceil \frac{3N}{2(k_\mathcal{C}+1)}\rceil \frac{\pi(k_\mathcal{C}+1)}{N} \right) }{ \sin\left( \lceil \frac{3N}{2(k_\mathcal{C}+1)}\rceil \frac{\pi}{N} \right) } \right\rbrace=0. $$
In numerical experiments we observe that this equation is fulfilled in a region where $\min\{\cdots\} = \sin\left(2\pi(k_\mathcal{C}+1)/N \right)/ \sin\left( 2\pi/N \right)$, when $N>11$. Thus, for $N>11$, the critical degree is given by the equation
\begin{equation}\label{eq_kc_kcycles}
    2 - \frac{ \sin\left(\pi(k_\mathcal{C}+1)/N\right) }{ \sin\left(\pi/N\right) } - \frac{ \sin\left(2\pi(k_\mathcal{C}+1)/N\right) }{ \sin\left(2\pi/N\right) } = 0,
\end{equation}
which in the thermodynamic limit holds
$$ \frac{\lambda_F[\mathcal{C}_N(k)]}{k} + \frac{\lambda_M[\mathcal{C}_N(k)]}{k} \to 2 - \textnormal{sinc}(\pi\,\rho_\mathcal{C}) - \textnormal{sinc}(2\pi\,\rho_\mathcal{C}) = 0. $$
Consequently, a solution for $\lambda_F/k_\mathcal{C} + \lambda_M/k_\mathcal{C}=2$ in the thermodynamic limit is $\rho_\mathcal{C}=1$, making $\lambda_F/k + \lambda_M/k < 2, \;\forall\,\rho\in(0,1)$, for infinite sized $k$-cycles.

%
   \subsection*{$k$-M{\"o}bius ladders -- Minimum and Maximum Laplacian Eigenvalues}\hypertarget{k-ML}{}
These graphs are defined by the Laplacian given in Eq.~(\ref{eq_kMobius_def}), implying that the eigenvalues $\lambda_n[\mathcal{M}_N(k)]$ (with $n = 0,\ldots,N-1$) from Eq.~(\ref{eq:eigenvals}) are
\begin{equation}\label{eq:eigenvals_kMobius_sup}
    \lambda_n[\mathcal{M}_N(k)] = k - 2\cos\left(\frac{2\pi n}{N}\right) - \sum_{j=(N+5-k)/2}^{(N-1+k)/2} \cos\left( \frac{2\pi\,n}{N} (j-1) \right)\!.
\end{equation}
Here we derive an explicit expression for the sum on the right-hand-side ($r.h.s.$) of Eq.~(\ref{eq:eigenvals_kMobius_sup}) by using complex exponentials and shifting the $j-1$ index to $j$. That is,
$$    r.h.s. = \sum_{j=(N+5-k)/2}^{(N-1+k)/2} \cos\left( \frac{2\pi\,n}{N} (j-1) \right) = \sum_{j=(N+3-k)/2}^{(N-3+k)/2} \cos\left( \frac{2\pi\,n}{N} j \right) = $$
$$  = \frac{1}{2} \sum_{j=(N+3-k)/2}^{(N-3+k)/2} \left\lbrace \exp\left[i\frac{2\pi\,n}{N}j\right] + \exp\left[-i\frac{2\pi\,n}{N}j\right] \right\rbrace, $$
where we shift $j$ again, such that $j' = j - (N+3-k)/2$; namely,
$$ r.h.s. = \frac{1}{2}\exp{ \left[ i\frac{\pi\,n}{N}\left(N-k+3\right) \right]} \sum_{j'=0}^{k-3} \exp{ \left[i\frac{2\pi\,n}{N}j'\right] } + $$
$$ + \frac{1}{2}\exp{ \left[ -i\frac{\pi\,n}{N}\left(N-k+3\right) \right] } \sum_{j'=0}^{k-3}\exp{ \left[-i\frac{2\pi n}{N}j'\right] }.$$
We then substitute the resultant geometric sums of $j'$ into $r.h.s.$,
$$ r.h.s. = \frac{1}{2}\exp{ \left[ i\frac{\pi\,n}{N}\left(N-k+3\right) \right] } \left( \frac{ 1 - \exp\left[i\frac{2\pi\,n}{N}\left(k-2\right)\right] }{ 1 - \exp\left[i\frac{2\pi\,n}{N}\right] } \right) + $$
$$ + \frac{1}{2}\exp{ \left[ -i\frac{\pi\,n}{N}\left(N-k+3\right) \right] } \left( \frac{1 - \exp{ \left[-i\frac{2\pi\,n}{N}\left(k-2\right)\right] } }{ 1 - \exp{ \left[-i\frac{2\pi n}{N}\right] } } \right), $$
which we can transform using the fact that $1 - \exp[\pm\,i\,\phi] = \pm2i\,\sin(\phi/2)\exp[\pm\,i\,\phi/2]$ for an arbitrary phase variable $\phi$. Starting by transforming the denominator and then the numerator of the geometric sums, we have
$$ r.h.s. = \frac{1}{4i}\exp{ \left[ i\frac{\pi\,n}{N}\left(N-(k-3)\right) \right] } \exp{ \left[ -i\frac{\pi\,n}{N} \right] } \left( \frac{ 1 - \exp\left[ i\frac{2\pi n}{N}\left(k-2\right)\right] }{\sin(n\,\pi/N)} \right) $$
$$ - \frac{1}{4i}\exp{ \left[ -i\frac{\pi\,n}{N}\left(N-(k-3)\right) \right] } \exp{ \left[i\frac{\pi\,n}{N} \right] } \left( \frac{1 - \exp\left[-i\frac{2\pi\,n}{N}\left(k-2\right)\right] }{ \sin(n\,\pi/N) } \right) = $$
$$ = \frac{2i}{4i}\exp{ \left[ i\frac{\pi\,n}{N}\left(N-(k-2)\right) \right] } \exp{ \left[ i\frac{\pi\,n}{N}\left(k-2\right) \right] }  \frac{ \sin\left(n\,\pi\,(k-2)/N\right) }{\sin(n\,\pi/N)} + $$
$$ + \frac{2i}{4i}\exp{ \left[ -i\frac{\pi\,n}{N}\left(N-(k-2)\right) \right] } \exp{ \left[-i\frac{\pi\,n}{N}\left(k-2\right) \right] } \frac{ \sin\left(n\,\pi\,(k-2)/N\right) }{ \sin(n\,\pi/N) } = $$
$$ = \frac{1}{2}\exp{ \left[ i\frac{\pi\,n}{N}N \right] } \frac{ \sin\left(n\,\pi\,(k-2)/N\right) }{\sin(n\,\pi/N)} + \frac{1}{2}\exp{ \left[ -i\frac{\pi\,n}{N}N \right] } \frac{ \sin\left(n\,\pi\,(k-2)/N\right) }{\sin(n\,\pi/N)}\;\Rightarrow $$
$$ \Rightarrow\; r.h.s. = \cos\left(n\,\pi\right) \frac{ \sin\left(n\,\pi\,(k-2)/N\right) }{\sin(n\,\pi/N)} = (-1)^n\frac{ \sin\left(n\,\pi\,(k-2)/N\right) }{\sin(n\,\pi/N)}. $$
Consequently, our explicit expression for Eq.~(\ref{eq:eigenvals_kMobius_sup}) is
\begin{equation}
    \lambda_n[\mathcal{M}_N(k)] = k - 2\cos\left(\frac{2\pi\,n}{N}\right) - (-1)^n \frac{ \sin\left(n\,\pi\,(k-2)/N\right) }{ \sin(n\,\pi/N) },
\end{equation}
which shows the ring contribution (first two terms) and the rungs (last term). In particular, using that $2\cos(\beta)\sin(\alpha) = \sin(\alpha+\beta) + \sin(\alpha-\beta)$, we get
\begin{equation}
    \lambda_n[\mathcal{M}_N(k)] = k + 1 - \left[ \frac{ \sin\left(3\pi\,n/N\right) + (-1)^n \sin\left(n\,\pi\,(k-2)/N\right) }{ \sin(n\,\pi/N) } \right]\!,
    \label{eq_kMobius_SIexp}
\end{equation}
where we note similarities (the term $k + 1$, the facts that $\lambda_n[\mathcal{M}_N(k)] = 0$ for $n = 0$ and $\lambda_n[\mathcal{M}_N(N-1)] = N-1$ for $n > 0$, and the symmetry in $n$ around $N/2$) and differences (terms withing brackets) to Eq.~(\ref{eq_kcycle_SIexp}) for $k$-cycles.

Here, $\lambda_1[\mathcal{M}_N(k)]$ is no longer the Fiedler eigenvalue -- as in Eq.~(\ref{eq_kCycleMinEig}) for $k$-cycles -- but the maximum eigenvalue, for almost any $k$. In order to show this, we note that the denominator in the bracketed expression is a monotonically increasing function of $n\in(0,N/2]$, meaning that the bracketed term becomes more significant the smaller the $n$. A negative numerator always tends to maximise the eigenvalue, which happens when $n$ is odd as long as both sines in the numerator do not change sign. In particular, the numerator is negative for $n = 1$, $\sin\left(3\pi/N\right) - \sin\left(\pi\,(k-2)/N\right) < 0$, as long as $k\in(5,N-1)$. However, as $n$ is increased, the denominator increases as well, decreasing the contribution from the bracketed term. As a result,
\begin{equation}
    \lambda_M[\mathcal{M}_N(k)] \equiv \max_{n\in[1,N/2]}\{\lambda_n[\mathcal{M}_N(k)]\} = \lambda_{1}\;\;\textnormal{if}\;k\in(6,N-1).
    \label{eq_kMobiusMaxEig}
\end{equation}
For $k \leq 6$, numerical experiments can be performed to find the eigenmode that maximises the Laplacian eigenvalue in Eq.~(\ref{eq_kMobius_SIexp}). For example, when $k = 6$, we find that $k$-M{\"o}bius ladders with $N$ even have a maximum eigenvalue that is approximately the mode $n/N\simeq 0.412$; and when $k = 3$, the maximum eigenvalue is given by $n/N = 0.5$, which means that $\lambda_M[\mathcal{M}_N(3)] = \lambda_{N/2}$. This shows that the mode of the maximum eigenvalue for $k$-M{\"o}bius ladders changes according to the network size and degree when $k \leq 6$.


Now, we argue that the Fiedler eigenvalue $\lambda_F[\mathcal{M}_N(k)]$ corresponds to the next lower eigenmodes. In particular, we find that
\begin{equation}
    \lambda_F[\mathcal{M}_N(k)] \equiv \min_{n\in[1,N/2]}\{\lambda_n[\mathcal{M}_N(k)]\} = \lambda_2\;\;\textnormal{if}\;k\in[3,k_c],
    \label{eq_kMobiusMinEig}
\end{equation}
where $k_c < N/2$ is derived from the transcendental identity $\lambda_2 = \lambda_3$, which is when the Fiedler becomes the third eigenmode instead of the second. Explicitly,
$$ \left[ \frac{ \sin\left(6\pi/N\right) +  \sin\left(2\pi\,(k_c-2)/N\right) }{ \sin(2\pi/N) } \right] = \left[ \frac{ \sin\left(9\pi/N\right) - \sin\left(3\pi\,(k_c-2)/N\right) }{ \sin(3\pi/N) } \right], $$
which approximately holds $k_c \simeq (2N+8)/5$. Similarly, we find that
\begin{equation}
    \lambda_F[\mathcal{M}_N(k)] \equiv \min_{n\in[1,N/2]}\{\lambda_n[\mathcal{M}_N(k)]\} = \lambda_3\;\;\textnormal{if}\;k\in(k_c,k_c'],
    \label{eq_kMobiusMinEig2}
\end{equation}
where $k_c'$ is derived from the transcendental identity $\lambda_3 = \lambda_4$, which reads
$$ \left[ \frac{ \sin\left(9\pi/N\right) -  \sin\left(3\pi\,(k_c-2)/N\right) }{ \sin(3\pi/N) } \right] = \left[ \frac{ \sin\left(12\pi/N\right) + \sin\left(4\pi\,(k_c-2)/N\right) }{ \sin(4\pi/N) } \right]. $$
Other critical degrees follow, progressively increasing the eigenmode that corresponds to the Fidler eigenvalue until converging to the complete graph, where $k = N-1$ and all eigenvalues are the same and hold $\lambda_n[\mathcal{M}_N(k=N-1)] = N,\;\forall\,n>0$.

Considering Eqs.~(\ref{eq_kMobiusMaxEig}) and (\ref{eq_kMobiusMinEig}), we have that, for $k\in(6,k_c]$,
$$ \frac{\lambda_M}{k} + \frac{\lambda_F}{k} = \frac{\lambda_1[\mathcal{M}_N(k)]}{k} + \frac{\lambda_2[\mathcal{M}_N(k)]}{k} = 2\frac{(k+1)}{k},
$$
$$ -\frac{1}{k} \left[ \frac{ \sin\left(3\pi/N\right) - \sin\left(\pi\,(k-2)/N\right) }{ \sin(\pi/N) } \right]  - \frac{1}{k}\left[ \frac{ \sin\left(6\pi/N\right) + \sin\left(2\pi\,(k-2)/N\right) }{ \sin(2\pi/N) } \right]. $$
This equation has two solutions: $\lambda_2/k + \lambda_1/k < 2$ when $7 \leq k < k_{\mathcal{M}}$ and $\lambda_2/k + \lambda_1/k > 2$ when $ k_{\mathcal{M}} < k \leq k_c \simeq (2N+8)/5$, $k_{\mathcal{M}}$ being the critical degree determined by the case when $\lambda_F/k_{\mathcal{M}} + \lambda_M/k_{\mathcal{M}} = 2$. That is,
$$ \frac{(k_\mathcal{M}+1)}{k_\mathcal{M}} - 1 = \frac{ \sin\left(3\pi/N\right) - \sin\left(\pi\,(k_\mathcal{M}-2)/N\right) }{ 2k_\mathcal{M}\sin(\pi/N) } + \frac{ \sin\left(6\pi/N\right) + \sin\left(2\pi\,(k_\mathcal{M}-2)/N\right) }{ 2k_\mathcal{M}\sin(2\pi/N) },$$
$$ 1 = \left[\frac{\sin\left(2\pi\,(k_\mathcal{M}-2)/N\right)}{2\sin(2\pi/N)} - \frac{\sin\left(\pi\,(k_\mathcal{M}-2)/N\right)}{2\sin(\pi/N)} \right] + \frac{ \sin\left(3\pi/N\right) }{ 2\sin(\pi/N) } + \frac{ \sin\left(6\pi/N\right) }{ 2\sin(2\pi/N) }, $$
\begin{equation}
    \alpha_N = \frac{ \sin\left(2\pi\,(k_\mathcal{M}-2)/N\right) }{ \sin(2\pi/N) } - \frac{ \sin\left(\pi\,(k_\mathcal{M}-2)/N\right) }{ \sin(\pi/N) },
    \label{eq_CriticDegree}
\end{equation}
where we define a constant, $\alpha_N \equiv 2 - \sin\left(3\pi/N\right)/\sin(\pi/N) - \sin\left(6\pi/N\right)/\sin(2\pi/N)$, which solely depends on $N$. Thus, Eq.~(\ref{eq_CriticDegree}) is a transcendental equation that allows to determines the critical degree that differentiates between $2$ classes of $k$-M{\"o}bius ladders: those such that $\lambda_F/k_{\mathcal{M}} + \lambda_M/k_{\mathcal{M}} < 2$ and those that $\lambda_F/k_{\mathcal{M}} + \lambda_M/k_{\mathcal{M}} > 2$.

We note that when $N\to\infty$, we can use Eqs.~(\ref{eq:lamda_M_mob}) and (\ref{eq:lamda_F_mob}) in the thermodynamic limit of $\lambda_1[\mathcal{M}_N(k)]$ and $\lambda_2[\mathcal{M}_N(k)]$. As a result, we get
\begin{equation}
    \frac{\lambda_1[\mathcal{M}_N(k)]}{k} + \frac{\lambda_2[\mathcal{M}_N(k)]}{k} \to 2 + \textnormal{sinc}(\pi\,\rho) - \textnormal{sinc}(2\pi\,\rho).
    \label{eq_InfCriticDensit}
\end{equation}
Consequently, there is a critical link density for infinite-sized $k$-M{\"o}bius ladders, $\rho_\mathcal{M}$, when $\textnormal{sinc}(\pi\,\rho_\mathcal{M}) - \textnormal{sinc}(2\pi\,\rho_\mathcal{M}) = 0$, with the solutions $\rho_\mathcal{M} = 0$ and $\rho_\mathcal{M} = 1$. This means that infinite-sized $k$-M{\"o}bius ladders fulfill $\lambda_2[\mathcal{M}_\infty(\rho)]/k + \lambda_1[\mathcal{M}_\infty(\rho)]/k > 2$, valid for $\rho\in(0,1)$, and coincide with the $k$-cycles on the complete graphs for $\rho = 1$.

%
\section*{Acknowledgements}
J.G. acknowledges funds from the Agencia Nacional de Investigaci{\'o}n e Innonvaci{\'o}n (ANII), Uruguay, POS$\_$NAC$\_$2018$\_$1$\_$151185, and the Comisi{\'o}n Academica de Posgrado (CAP), Universidad de la Rep{\'u}blica, Uruguay. Both authors acknowledge funds from the Comision Sectorial de Investigaci{\'o}n Cientif{\'i}ca (CSIC), Uruguay, group grant ``CSIC2018 - FID13 - grupo ID 722''.

\section*{Author Contributions}
{\bf Juan Gancio:} Formal analysis, Visualization, Writing - Original Draft. {\bf Nicolás Rubido:} Conceptualization, Visualization, Writing - Review \& Editing, Supervision.


\newpage
\end{document}